\documentclass{aa}

\usepackage{graphicx}
\usepackage{txfonts}
\usepackage{aalongtable}
\usepackage{natbib}
\bibpunct{(}{)}{;}{a}{}{,}
\begin{document}
   \title{A Study of Catalogued Nearby Galaxy Clusters in the SDSS-DR4}

   \subtitle{I. Cluster Global Properties}

   \author{J. A. L. Aguerri, R. S\'anchez-Janssen \& C. Mu\~noz-Tu\~n\'on}

   \offprints{J. A. L. Aguerri}

   \institute{Instituto de Astrof\'{\i}sica de Canarias
              C/ V\'{\i}a L\'actea s/n, 38200 La Laguna, Spain.
              \email{jalfonso@iac.es, ruben@iac.es, cmt@iac.es}}

   \date{Received ; accepted }

   \authorrunning{Aguerri et al.}
   \titlerunning{Global Properties of Nearby Galaxy Clusters}

 
  \abstract
   {Large surveys as the Sloan Digital Sky Survey have made available  large amounts of spectroscopic and photometric data of galaxies, providing important information for the study of galaxy evolution in dense environments.}
   {We have selected a sample of 88 nearby ($\rm{z}<0.1$) galaxy clusters from the SDSS-DR4 with redshift information for the cluster members. In particular, we focus our results on the galaxy morphological distribution, the velocity dispersion profiles, and the fraction of blue galaxies in clusters.}
   {Cluster membership was determined using the available velocity information. We have derived global properties for each cluster, such as their mean recessional velocity, velocity dispersion, and virial radii. Cluster galaxies have been grouped in two families according to their $u-r$ colours.}
   {The total sample consists of 10865 galaxies. As expected, the highest fraction of galaxies (62$\%$) turned to be early-type (red) ones, being located at smaller distances from the cluster centre and showing lower velocity dispersions than late-type (blue) ones. The brightest cluster galaxies are located in the innermost regions and show the smallest velocity dispersions. Early-type galaxies also show constant velocity dispersion profiles inside the virial radius and a mild decline in the outermost regions. In contrast, late-type galaxies show always decreasing velocity dispersions profiles. No correlation has been found between the fraction of blue galaxies and cluster global properties,such as cluster velocity dispersion and galaxy concentration. In contrast, we found correlation between the X-ray luminosity and the fraction of blue galaxies.}
   {These results indicate that early- and late-type galaxies may have had different evolution. Thus, blue galaxies are located in more anisotropic and radial orbits than early-type ones. {\bf{Their star formation seems to be independent of the cluster global properties in low mass clusters, but not for the most massive ones.}} We consider that it is unlikely that the whole blue population consists of recent arrivals to the cluster. {\bf{These observational results suggest that the global environment could be important for driving the evolution of galaxies in the most massive cluster ($\sigma > 800$ km s$^{-1}$). However, the local environment could play a key role in galaxy evolution for low mass clusters.}}}


   \maketitle
%

\section{Introduction}

The large amount of spectroscopic and photometric data obtained during the last years by surveys such as the Sloan Digital Sky Survey \citep[SDSS; ][]{york00} or the 2dF Galaxy Redshift Survey \citep[2dFGRS; ][]{colless01} have opened a new horizon for the study of galaxy evolution, and in particular in the study of galaxy clusters. It is well known that the environment plays an important role in the evolution of galaxies, and it is one of the keys that a good galaxy evolution theory should address. There are several physical mechanisms, not present in the field, which can dramatically transform  galaxies in high density environments. Galaxies in clusters can evolve due to, e.g., dynamical friction, which can slow down the more massive galaxies, circularise their orbits and enhance their merger rate \citep{denhartog96, mamon92}. Interactions with other galaxies and with the cluster gravitational potential can disrupt the outermost regions of the galaxies and produce galaxy morphological transformations from late- to early-types \citep{moore96}, or even change massive galaxies into dwarf ones \citep{mastroprieto05}. Swept of cold gas produced by ram pressure stripping \citep{gunn72, quilis00} or swept of the hot gas reservoirs \citep{bekki02} can alter the star formation rate (SFR) of galaxies in clusters. But it is still a matter of debate which of these mechanisms is the main responsible of the galaxy evolution in galaxy clusters \citep[see ][]{goto05}. Nevertheless, it is clear that all of these mechanisms transform  galaxies from late- to early-types, and can produce the different segregations observed in galaxy clusters. 

One of the first segregations discovered in galaxy clusters was the morphological one. The first evidences of such segregation date from Curtis (1918) and Hubble \& Humason (1931), and was quantified by Oemler (1974) and Melnick \& Sargent (1977). In a thorough work, Dressler (1980) analysed a sample of 55 nearby galaxy clusters, containing over 6000 galaxies, and observed that elliptical and S0 galaxies represent the largest fraction of galaxies located in the innermost and denser regions of galaxy clusters. In contrast, the outskirts of the clusters were dominated by spiral galaxies. In more distant clusters the fraction of E galaxies is as large or larger than in low-redshift clusters, but the S0 fraction is smaller \citep[][]{dressler97, fasano00}. This has been interpreted as an evolution with redshift, being late-type galaxies transformed into early-type ones. Segregations in velocity space have also been observed in galaxy clusters. Early observations found that E and S0 galaxies showed smaller velocity dispersions than spirals and irregulars \citep[][]{tamman72, melnick77, moss77}. This has also been confirmed by other authors during the last two decades \citep[][]{sodre89, biviano92, andreon96, stein97}. The data from the ENACS survey \citep[][]{katgert98} produced a large sample of galaxies with spectroscopic redshifts and shed more light to this problem. Thus, Adami et al. (1998) studied a sample of 2000 galaxies, confirming early findings that the velocity dispersion of galaxies increases along the Hubble sequence: E/S0 galaxies show smaller velocity dispersions than early- and late-type spirals. This segregation was also observed in the velocity dispersion profiles (VDPs): late-type galaxies have  decreasing VDPs, while E, S0 and early spirals show almost flat VDPs \citep[][]{adami98}. The different kinematics shown by the different types of galaxies was analysed in more detail by Biviano \& Katgert (2004) who found that the velocity segregation of the different Hubble types is due to differences in orbits. Thus, early-type spirals have isotropic orbits, while late-type ones are located in more anisotropic orbits. The observed morphological and velocity segregation in clusters have been usually used  to conclude that late-type spiral galaxies in clusters are recent arrivals to the cluster potential \citep[][]{stein97, adami98}.



Star formation in galaxies is also affected by the environment. Butcher \& Oemler (1984) found that the fraction of blue galaxies, $f_{b}$, in clusters is smaller than in the field and evolves with redshift: more distant clusters show larger values of $f_{b}$. This trend was interpreted as an evolutionary effect of the SFR in galaxy clusters. But the significant increase of new data has made it clear that the Butcher-Oemler effect is not only an evolutionary trend. A large scatter in the values of $f_{b}$ has been observed in narrow redshift ranges \citep[][]{smail98, margoniner00, goto03}, which suggests that the variation of $f_{b}$ is influenced by environmental effects. In the past, many authors have tried to find  correlations of $f_{b}$ with cluster properties, such as X-ray luminosity \citep[][]{andreon99, smail98, fairley02}, luminosity limit and clustercentric distance \citep[][]{ellingson01, goto03, depropris04}, richness \citep[][]{margoniner01, depropris04}, cluster concentration \citep[][]{butcher84,depropris04}, presence of substructure \citep[][]{metevier00} or cluster velocity dispersion \citep[][]{depropris04}. Some of these works found correlations between $f_{b}$ and the cluster environment while others did not, being such connection still a matter of debate. However, these works were usually done using small and heterogeneous cluster samples \citep[but see e.g., ][]{depropris04}.

Environmental effects have also been invoked in order to explain the differences  between the photometrical components of cluster and field spiral galaxies. Thus, it has been observed that the scale-lengths of the disks of spiral galaxies in the Coma cluster  are smaller than those of similar galaxies in the field \citep[][]{gutierrez04, aguerri04}. Interactions between galaxies or with the cluster potential can disrupt the disks of spiral galaxies in clusters. They can be strong enough for transforming bright late-type spiral galaxies in dwarfs \citep[][]{aguerri05a}.  The disrupted material would be part of the intracluster light already detected in some nearby galaxy clusters \citep[][]{arnaboldi02, arnaboldi04, aguerri05b} and galaxy groups \citep[][]{castrorodriguez03, aguerri06}.

The observational results summarised before illustrate the important role played by environment in galaxy evolution. They also indicate that late-type and early-type galaxies in clusters are two different families of objects with different properties, which points to different origins or evolution. Nevertheless, the main mechanisms responsible of this different evolution still remain unknown. In the present paper, we study one of the largest and more homogeneous galaxy cluster sample available in the literature.  We have obtained the cluster membership, mean velocity, velocity dispersion, virial radius and positions for a sample of 88 clusters located at $z < 0.1$.  We have investigated the main properties of a large sample of early (red) and late (blue) types of galaxies, such as their location within the cluster, their mean velocity dispersion, their VDPs, the $L_{X}-\sigma$ relation, and the fraction of blue galaxies for each cluster. This work provides important information about the properties of galaxies in nearby clusters, which will be useful in order to put constraints on cosmological models of cluster formation. This is the first paper of a series in which we will analyse the properties of the dwarf galaxy population (S\'anchez-Janssen et al. in preparation), substructure in galaxy clusters (Aguerri et al. in preparation), and composite luminosity function of galaxy clusters (S\'anchez-Janssen et al., in preparation).

The paper is organised as follows. Section 2 shows the discussion about the galaxy cluster sample. The cluster membership and cluster global parameters are presented in Section 3. The results obtained about the morphological segregation, velocity dispersion profiles, L$_{X}-\sigma$ relation, and the fraction of blue galaxies are given in Sections 4, 5, 6 and 7, respectively. The discussion and conclusions are presented in Sections 8 and 9, respectively. Throughout this work we have used the cosmological parameters: $H_{o}=75$ km s$^{-1}$ Mpc$^{-1}$, $\Omega_{m}=0.3$ and $\Omega_{\Lambda}=0.7$.


\section{Galaxy cluster Sample}

We have used photometric and spectroscopic data of  objects classified as galaxies from the SDSS-DR4, an imaging and spectroscopic survey of a large area in the sky \citep[][]{york00}. The imaging survey was carried out through five broad-band filters, $u g r i z$, spanning the range from 3000 to 10000 $\AA$, reaching a limiting $r$-band magnitude $\approx22.2$ with 95$\%$ completeness, and covering an area of 6670 deg$^{2}$  \citep[][]{adelman06}. A series of pipelines process the imaging data and perfom the astrometric calibration \citep[][]{pier03}, the photometric reduction \citep[][]{lupton02} and the photometric calibration \citep[][]{hogg01}. Objects brighter than $m_{r}=17.77$ were selected as possible targets for the spectroscopic survey, covering an area of 4783 deg$^{2}$ of the sky for the DR4. The spectroscopic data were obtained with optical fibers with a diameter of 3$^{''}$ at the focal plane, resulting in an spectral covering in the wavelength range 3800--9200 $\AA$ with a resolution of $\lambda/\Delta \lambda \approx 2000$.

Our  sample consists of all clusters with known redshift at $z<0.1$ from the catalogues of Abell et al. (1989), Zwicky et al. (1961), B{\"o}hringer et al.(2000) and Voges et al. (1999) that have been mapped by the SDSS-DR4.  We downloaded only those galaxies located within a radius of 4.5 Mpc around the centres of the galaxy clusters. Only those clusters with more than 30 galaxies with spectroscopic data in the searching radius were considered, resulting in a sample formed by 240 clusters following the previous criteria.  The SDSS-DR4 spectroscopic galaxy target selection was done by an automatic algorithm \citep[see ][]{strauss02}. The main galaxy sample consists of galaxies with $r$-band Petrosian magnitudes brighter than 17.77 and $r$-band Petrosian half-light surface brightness brighter than 24.5 mag arcsec$^{-2}$. The completeness of this sample is high, exceeding $99\%$ \citep[see ][]{strauss02}. However, some of the selected spectroscopic targets were not observed at the end. This incompleteness has several causes, including the fact that two spectroscopic fibers cannot be placed closer than 55$^{''}$ on a given plate, possible gaps between the plates, fibers that fall out of their holes, and so on. According to these reasons, we expect that the incompleteness of the spectroscopic data will be more important for bright galaxies in high density environments such  as galaxy clusters. Figure 1 shows the mean completeness\footnote{We have defined the spectroscopic completeness per magnitude bin as the ratio of the number of galaxies with spectroscopic data to the number of galaxies with photometric information.} of the SDSS-DR4 spectroscopic data as a function of the $r$-band magnitude for the selected galaxies, where a fast increment towards faint magnitudes can be observed. In order to avoid possible effects on the results due to this effect, we have completed the spectroscopic SDSS-DR4 observations with the data available at the Nasa Extragalactic Database (NED). Figure 1 also shows the mean completeness as a function of $r$-band magnitude after the spectroscopic data from NED were included in the sample. Notice that the new mean completeness is almost constant ($\approx$ 85$\%$) for all magnitudes brighter than $m_{r}=17.77$.  We have made a second selection of the clusters by considering only those from our original list with completeness larger than 70$\%$ for galaxies brighter than 17.77 in the $r$-band.

   \begin{figure}
   \centering
   \includegraphics[width=9cm]{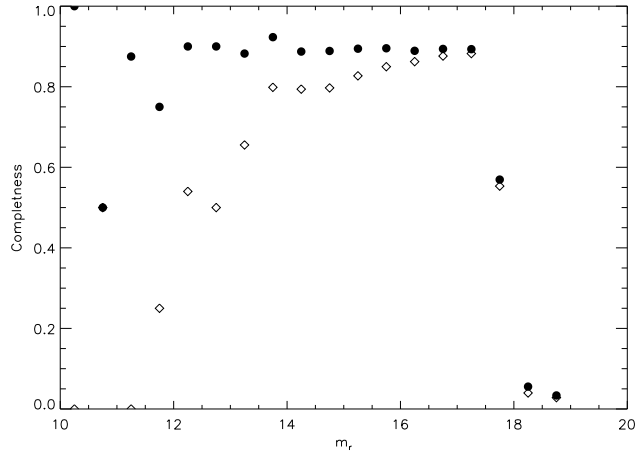}
      \caption{Mean completeness of the cluster sample as a function of the $r$-band magnitude. Diamonds represent the spectroscopic data from SDSS-DR4 and black circles after the completion with data from NED.
              }
   \end{figure}
%

\section{Cluster Membership}

Clusters properties such as the  mean cluster velocity, the velocity dispersion, the cluster centre or the virial radius can be significantly affected by projection effects.  Several methods have been developed during decades in order to obtain reliable galaxy cluster membership and avoid the presence of interlopers. They can be classified in two families. First, those algorithms that use only the information in the velocity space, e.g. 3$\sigma$-clipping techniques \citep[][]{yahil77}, gapping procedures \citep[][hereafter ZHG algorithm]{beers90, zabludoff90} or the KMM algorithm \citep[][]{ashman94}. The other family  corresponds to those algorithms which use information of both position and velocity, such as the methods designed by Fadda et al. (1996), den Hartog \& Katgert (1996), or Rines et al. (2003).  

The cluster membership in our sample was obtained using a combination of two algorithms. A first rough cluster membership determination was obtained using the ZHG algorithm, which in a second step  was then refined using the KMM algorithm. The ZHG algorithm is a typical gapping procedure which determines the cluster membership by the exclusion of those galaxies located at more than a certain velocity distance ($\Delta$v) from the nearest galaxy in the velocity space. Then, the mean velocity (v$_{m}$) and velocity dispersion ($\sigma$) of the remaining galaxies are calculated. After sorting objects with velocities greater than v$_{m}$, any galaxy separated in velocity more than $\sigma$ from the previous one is classified as non member. The same is done for those galaxies with velocities less than v$_{m}$. The process is repeated several times and finally the mean cluster velocity (v$_{c}$) and the cluster velocity dispersion ($\sigma_{c}$) are obtained. Zabludoff et al. (1990) pointed out that this method lacks statistical rigour and tends to give overestimated values of $\sigma_{c}$. One of the disadvantages of this method is that the results obtained strongly depend on the chosen value of $\Delta$v. Large values of $\Delta$v imply that a large fraction of interlopers are identified as cluster members. On the contrary, small values of $\Delta$v result in the lost of cluster galaxies. We have investigated the variation of $\sigma_{c}$ for different values of $\Delta$v, obtaining that $\Delta$v=500 km s$^{-1}$ is an appropriate value for our clusters. This method has also the advantage that has an easy implementation and does not require too much computational time.  Recently, it has been used in works involving a large number of clusters, such as those from the 2dFGRS \citep[][]{depropris03}. The ZHG algorithm splits the velocity histograms in different galaxy groups, being one of them located at the catalogued redshift of the cluster. That group was taken and analysed in more detail with the KMM algorithm. In the few cases where there was no galaxy group located at the catalogued redshift we identified the most significant groups having $z<0.1$ as the cluster itself.

The KMM algorithm \citep[][]{ashman94} estimates the statistical significance of bi-modality in a dataset. We have run it to the group of galaxies given by the ZHG algorithm which contains the catalogued redshift of the cluster. The KMM algorithm gives us the compatibility of the velocity distribution of such group of galaxies with a single  or multiple Gaussian distribution. We considered three different cases which are summarised in Fig. 2:

\begin{itemize}
\item Single cluster: the velocity distribution of the galaxies is compatible with a single Gaussian, e.g. Abell 757. 
\item Cluster with substructure: the velocity distribution is compatible with multiple groups. We identified the cluster as the group with the largest number of galaxies plus those groups which mean velocities lie within 3$\sigma$ from the mean velocity of the largest one\footnote{In this case $\sigma$ is the velocity dispersion of the largest group of galaxies.}, e.g. Abell 1003.
\item Cluster with contamination: the velocity distribution is  compatible with the presence of several groups, but the mean velocities of the smaller groups deviate more than 3$\sigma$ from the most populated one, which we identify as the cluster itself, e.g. Abell 168 .
\end{itemize}

We have explored the differences in the values of v$_{c}$ and $\sigma_{c}$ if we consider as interlopers those groups of galaxies located at a velocity distance larger than 1$\sigma$ or 3$\sigma$ from the mean velocity of the main galaxy group. We obtained that the differences in v$_{c}$ and $\sigma_{c}$ in 90$\%$ of the clusters are less than 20$\%$. The remaining 10$\%$ of the clusters are those with significant structure in the velocity distribution, being most of them more than one cluster along the line of sight. Thus, we have adopted $3\sigma$ as the default except for those clusters with significant differences between 1$\sigma$ and $3\sigma$, for which we have measured the mean velocity and velocity dispersion of the cluster adopting the criteria of 1$\sigma$. Through all of this process, the determination of v$_{c}$ and $\sigma_{c}$ was done using the biweight robust estimator of Beers et al. (1990).

   \begin{figure}
   \centering
   \includegraphics[width=9cm,height=10cm]{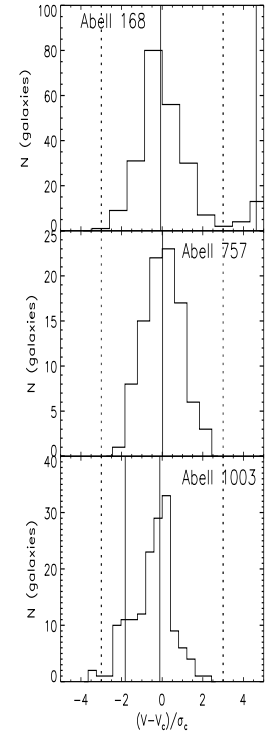}
      \caption{Velocity histograms of three representative clusters of the sample. The vertical full lines represent the mean velocity of the different groups of galaxies in which KMM algorithm has divided the velocity histogram. The dotted vertical lines represent v$_{c}\pm3\sigma_{c}$.
              }
   \end{figure}
%

\subsection{Cluster global parameters}

Once the cluster membership was determined, we obtained the global parameters of each cluster, i.e., mean velocity (v$_{c}$), velocity dispersion ($\sigma_{c}$), cluster centre, and the radius $r_{200}$. All of these parameters were computed using only the cluster members.

The determination of the cluster centre is important in order to accurately compute the other parameters of the clusters. The centre of the cluster is determined by the potential well, which can be traced by the position of the peak of the X-ray luminosity of the cluster. That peak was considered as the centre of those clusters from our sample with X-ray measurements in the literature. Unfortunately, not all the clusters from the sample have X-ray data. In that case, the centre of these clusters was determined by the peak of the galaxy surface density\footnote{The galaxy surface density was computed using the algorithm designed by Pisani (1996).}. For those clusters with X-ray data we have compared the centres given by the peaks of X-ray luminosity and galaxy surface density, obtaining a mean difference of 150 kpc.

 Analytic models \citep[][]{gott72} and simulations \citep[][]{cole96} indicate that the virialized mass of clusters is generally contained inside the surface where the mean inner density is 200$\rho_{c}$, where $\rho_{c}$ is the critical density of the Universe. The radius of that surface is called $r_{200}$.  We have computed the $r_{200}$ for our clusters using the same approximation as Carlberg et al. (1997): 
\begin{equation}
r_{200}=\frac{\sqrt{3} \sigma_{c}}{10~H(z_{c})},
\end{equation}

where $H(z_{c})$ is the Hubble constant at  the cluster redshift $z_{c}$.

The previous global parameters of the clusters (v$_{c}$, $\sigma_{c}$, $r_{200}$ and centre) were obtained as described above but in a recurrent way. In a first step, they where determined using all cluster member galaxies around 4.5 Mpc from the centre of the cluster. After this step we recalculated the parameters using only those galaxies located inside $r_{200}$. The method was repeated several times until the difference in the parameters obtained in two consecutive steps was less than 5$\%$. Three or four iterations were usually enough for reaching the convergence. In order to obtain reliable parameters of the clusters, those with less than 15 galaxies within $r_{200}$ were removed from our list. This results in a final sample formed by 110 nearby galaxy clusters. Table 1 shows the sample of galaxy clusters and their global parameters. The columns of Table 1 represent: (1) galaxy cluster name, (2, 3) cluster centres ($\alpha$ (J2000), $\delta$ (J2000)), (4) mean radial velocity, (5) cluster velocity dispersion, (6) $r_{200}$ radius, (7) number of galaxies within $r_{200}$, and (8) spectroscopic completeness.

For 6 clusters (Abell 1003, Abell 1032, Abell 1459, Abell 2023, Abell 2241 and ZwCl1316.4-0044) large differences in the mean recessional velocity have been found between the values given in Table 2 and those from NED. These are the clusters with no significant galaxy group at the catalogued redshift (see Section 3).

In order to consider the possible influence of neighbouring clusters on the global properties of the sample we searched in the surroundings of each cluster for the presence of companions. Following Biviano \& Girardi (2003), we have considered that two clusters, $i$ and $j$, are in interaction when:

\begin{equation}
|\rm{v}_{i}-\rm{v}_{j}|<3(\sigma_{i}+\sigma_{j}) \\
R_{i,j}<2(r_{200,i}+r_{200,j}),
\end{equation}

where R$_{i,j}$ is the projected distance between the centres of the clusters and v$_{i,j}$, $\sigma_{i,j}$, r$_{200,i,j}$ their respective mean velocities, velocity dispersions and $r_{200}$. We found 16 couples of clusters in interaction according to the previous criteria. The remaining sample (88 clusters) followed the isolation criteria, and will be used in the analysis presented in the following sections. Figure 3 shows the sky distribution of the cluster members and the galaxy velocities as a function of clustercentric distance for a sample of 8 clusters. Red points represent the galaxies taken as cluster members while black points are interlopers. Notice the large number of interlopers in some of the galaxy clusters, such as Abell 1291, Abell 1383, Abell 2244. Some of them, Abell 1291 and Abell 1383, were not included in the final isolated sample due to the presence of companions.

\subsection{Corrections to line-of-sight velocities}

Line-of-sight velocities of galaxies in clusters were corrected by two effects: cosmological redshift and global velocity field. We should take into account that we will compare the velocity dispersion of clusters located at different redshifts. Thus, for each galaxy we have $1+z_{obs}=(1+z_{c})(1+z_{gal})$ \citep[][]{danese80}, being $z_{obs}$ the apparent redshift of the galaxies, $z_{c}$ the cosmological redshift of the cluster, and $z_{gal}$ the redshift of the galaxy respect to the cluster centre. This correction can affect up to 10$\%$ for the most distant clusters in our sample.

Galaxy clusters are frequently part of larger cosmological structures such as filaments, superclusters or multiple systems, which can affect the velocity field resulting in a modified cluster velocity dispersion. The interaction between galaxy clusters can also produce distorted velocity fields. We have investigated the importance of these effects in the velocity field of our clusters by making a least-square fit to the radial velocities of  cluster galaxies with respect to their position in the plane of the sky \citep[see ][]{denhartog96, girardi96}. For each fit we computed the coefficient of multiple determination, $R^{2}$. In order to test the significance of the fitted velocity gradients, we run 1000 Monte Carlo simulations for each cluster for which the correlation between position and velocity was removed. This was achieved by shuffling the velocities of the galaxies with respect to their positions. We defined the significance of velocity gradients as the fraction of Monte Carlo simulations with $R^{2}$ smaller than the observed one. This correction of the velocity field was applied to those cluster in which the significance of velocity gradients is larger than 99$\%$ ($~$ 30\% of the total sample). However, this correction has small effects both in the shape of the velocity dispersion profiles and on the total velocity dispersion (the mean absolute correction was about $40\pm15$ km s$^{-1}$). This is in agreement with similar corrections applied in other cluster samples \citep[][]{denhartog96, girardi96}. 

   \begin{figure*}
   \centering
   \includegraphics[height=13cm]{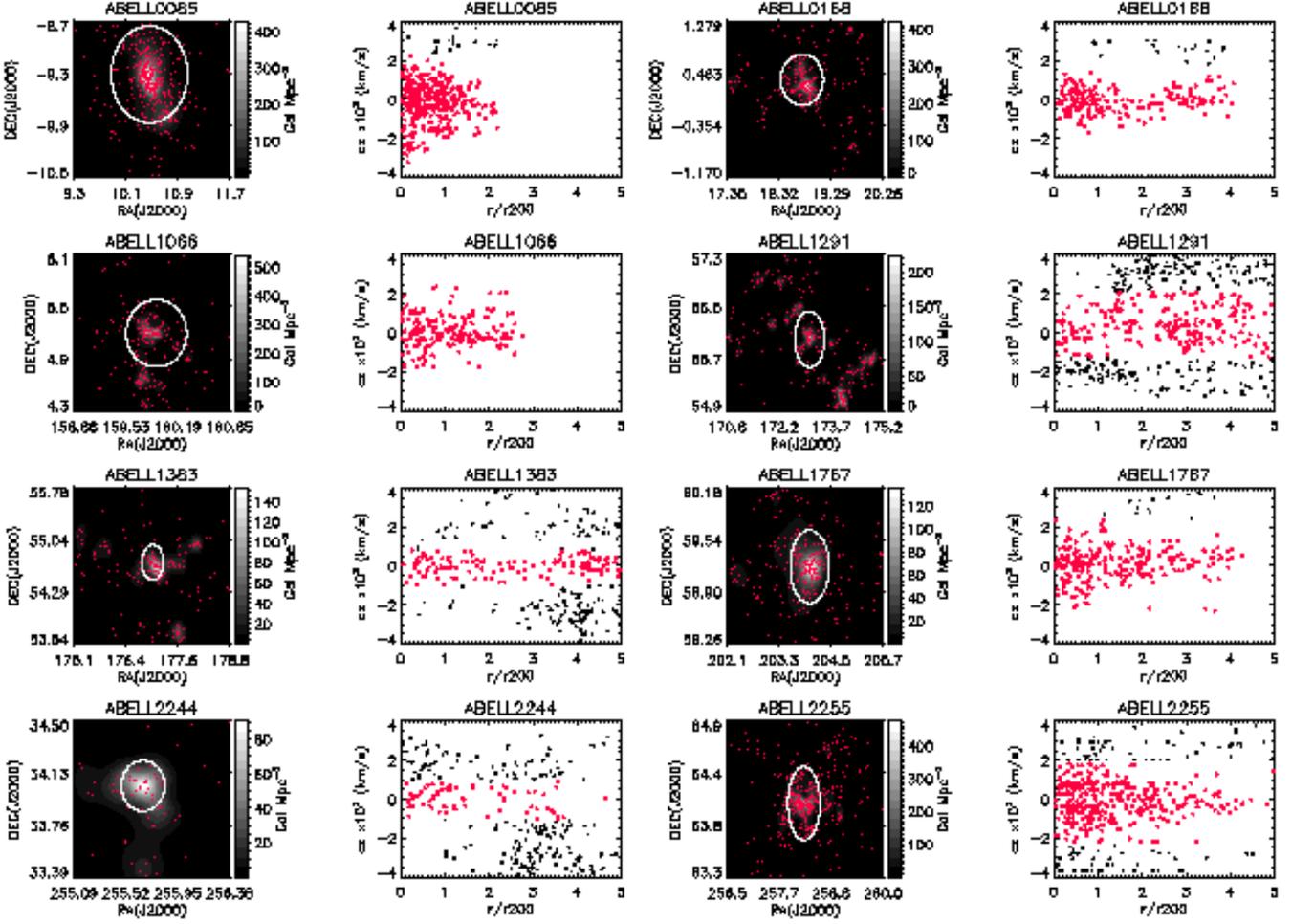}
      \caption{Galaxy surface density (images) and radial velocity versus distance to the cluster center for the galaxy cluster member (red points) of a subsample of 8 clusters. The overplotted circle have a radius equal to $r_{200}$ for each cluster. The black points represent interloper galaxies.}
   \end{figure*}
%

\subsection{Comparison with other methods}

Some of the clusters presented in our sample have been previously studied by other authors. However, we have avoided comparing our results with those from the literature given the different datasets used. In order to compare our cluster membership method with others proposed in the literature, we have computed $\sigma_{c}$ of our clusters with two more methods: a 3$\sigma$-clipping and the method proposed by Fadda et al. (1996). The median absolute difference between our $\sigma_{c}$ and those computed by the 3$\sigma$-clipping method is only 17 km s$^{-1}$. Only 10$\%$ of the clusters show important diferences ($\Delta \sigma_{c} > 200$ km s$^{-1}$) in the computation of the velocity dispersion of the cluster with the two methods. They correspond to those clusters affected by large amount of structure along the line of sight. The 3$\sigma$-clipping method gives for these clusters considerably larger values of $\sigma_{c}$ than ours. Differences were larger when we compared with Fadda's method. In this case the mean absolute difference in $\sigma_{c}$ between the two methods was 84 km s$^{-1}$ and 80$\%$ of the clusters show differences smaller than 200 km s$^{-1}$.

Recently, Popesso et al. (2006) have obtained the values of $\sigma_{c}$ for a sample of Abell clusters using SDSS-DR4 data, for which cluster membership was obtained using the selection algorithm of Katgert et al. (2004). The median absolute difference between our and their $\sigma_{c}$ is 45 km s$^{-1}$ for the 28 clusters in common. Only for 4 clusters (Abell 1750, Abell 1773, Abell 2244 and Abell 2255) the absolute differences in $\sigma_{c}$ is larger than 200 km s$^{-1}$. 

We have also compared our results with those given in the cluster catalogue presented by Miller et al. (2005). We found 16 clusters in common, being 74 km $s^{-1}$ the median absolute difference between our and their $\sigma_{c}$. In this case, 3 clusters show an absolute differences in $\sigma_{c}$ larger than 200 km s$^{-1}$.   

We can conclude that in most of the cases our cluster membership method reported values of $\sigma_{c}$ similar to those given by other methods. Only for 10-20$\%$ of the clusters the absolute differences in $\sigma_{c}$ between our method and the others is larger than 200 km s$^{-1}$. For these clusters the structure along the line of sight is the responsible of the difference, being our $\sigma_{c}$ values smaller than the others. 

\subsection{L$_{x}$-$\sigma$ relation}

We can learn about the nature of cluster assembly by studying the relations between cluster observables. One of the most universals is the well known relation between the cluster X-ray luminosity and the velocity dispersion of its galaxies ($L_{X} \propto \sigma_{c}^{b}$). Cluster formation models predict that if the only energy source in the cluster comes from the gravitational collapse, then $b\approx 4$. This relation has been studied in the literature by many authors using different cluster samples, finding values of $b$ between 2.9 and 5.3 \citep[][]{edge91, quintana82, mulchaey98, mahdavi01, girardi01, borgani99, xue00, ortizgil04, hilton05}. The study of the $L_{X}-\sigma_{c}$ relation in our cluster sample will be also useful as another check for the values of $\sigma_{c}$ we have derived. We have X-ray data for 48 galaxy clusters from our sample. The X-ray data have been obtained from Ebeling et al. (1998), B{\"o}hringer et al.(2000), Ebeling et al.(2000) and Ledlow et al.(2003), and the X-ray luminosities are measured in the ROSAT band (0.1-2.4 keV). 

Figure 4 shows the $L_{X}-\sigma$ relation for this subset with available X-ray data in the literature. The Spearman coefficient of the relation is 0.56 and the significance from zero correlation is greater than 3$\sigma$. This indicates the existence of a correlation between $L_{X}$ and $\sigma_{c}$ for the clusters of our sample.  We used the bivariate correlated errors and intrinsic scatter (BCES) bisector method of Akritas \& Bershady (1996) to obtain the coefficient and power-law slope estimates of the relation. This fitting technique takes into account errors in both variables and intrinsic scatter. The $L_{X}-\sigma_{c}$ relation for our clusters is given by:

\begin{equation}
L_{X}(0.1-2.4 {\rm ~keV})=10^{33.7\pm1.2} \sigma^{3.9\pm0.4}
\end{equation}

This result is in very good agreement with another measurement of this relation using the same ROSAT band (0.1-2.4 keV) for the X-ray data and the same fitting algorithm \citep[see ][]{hilton05}.

   \begin{figure}
   \centering
   \includegraphics[width=9cm]{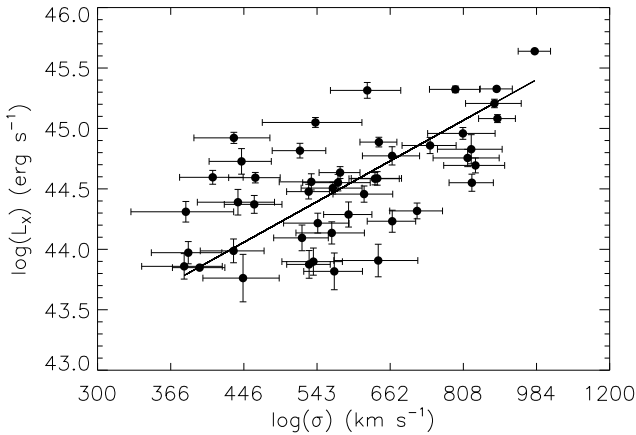}
      \caption{$L_{X}-\sigma$ relation for the 48 galaxy clusters with X-ray data in the ROSAT band (0.1-2.4 keV) from our sample. The full line represents the best fit using the BCES bisector algorithm (see text for more details).
              }
   \end{figure}
%

\subsection{Redshift distribution and sample completeness}

The 88 isolated galaxy clusters are located in a redshift range between 0.02 and 0.1, with an average redshift of 0.071. Figure 5 shows the absolute $r$-band magnitude ($M_{r}$) as a function of the redshift for the galaxies in our cluster sample\footnote{See section 3 for the explanation of the computation of the absolute magnitudes of the galaxies.}. It is clear that the completeness magnitude is a function of redshift. This figure shows that the full sample is complete for galaxies brighter than $M_{r}=-20.0$. The lack of completeness for fainter galaxies will be taken into account in the subsequent analysis.

   \begin{figure}
   \centering
   \includegraphics[width=9cm]{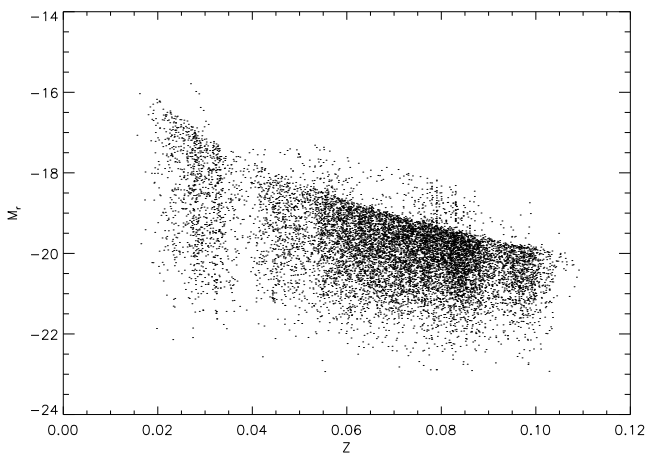}
      \caption{Absolute $r$-band magnitude as a function of redshift for the galaxies of our cluster sample.
              }
   \end{figure}
%

\section{Morphological Segregation}

Light concentration or colours have been used extensively in the literature in order to classify galaxies. Shimasaku et al. (2001) and Strateva et al. (2001) using SDSS data, found that the ratio of Petrosian 50 percent light radius to Petrosian 90 percent light radius, $C_{in}$, measured in the $r$-band image was a useful index for quantifying  galaxy morphology. Strateva et al. (2001) also found that the colour $u-r =$ -2.22 efficiently separates early- and late-type galaxies at $z < 0.4$. We have used colours for  classifying galaxies, because properties such as velocity dispersion in galaxy clusters are better correlated with galaxy colours than galaxy morphology \citep[][]{goto05}. The magnitude of the galaxies were corrected by two effects: Galactic absorption and $k-$correction. The Galactic absorption in the different filters was obtained from the dust maps of Schlegel et al. (1998). We applied the $k-$correction using the kcorrect.v4\_1\_4 code by Blanton et al. (2003) in order to obtain the rest-frame magnitudes of the galaxies for the different bandpasses. Once these two corrections were done, we classified the galaxies in red ($u-r \ge 2.22$) and blue ones ($u-r<2.22$).

The galaxy data was downloaded from the SDSS database according to a metric criteria: we downloaded the information of all galaxies located within a radius of 4.5 Mpc at each galaxy cluster redshift. This means that we are mapping different physical regions for each cluster. In order to avoid this problem we have studied the ratio $r_{max}/r_{200}$ for each cluster, being $r_{max}$ the maximum distance of a galaxy from the cluster centre for each galaxy cluster. We have obtained that all clusters of our sample reach $\frac{r_{max}}{r_{200}}=2$, and 50$\%$ of them reach $\frac{r_{max}}{r_{200}}=5$.  

Our sample of galaxies consists of 6880 galaxies located within a radius $2\times r_{200}$, being $62\%$ of them red galaxies and $38\%$ blue ones. If we consider all galaxies within $5\times r_{200}$ then the sample has 10865 galaxies, being 55$\%$ and 45$\%$ red and blue galaxies, respectively.  The red and blue galaxies were also grouped in three categories according to their $r$-band magnitude: $M_{r}<M_{r}^{*}-1$, $M_{r}^{*}-1<M_{r}<M_{r}^{*}+1$, and $M_{r}>M_{r}^{*}+1$\footnote{$M^{*}_{r}-5 log(h)=-20.04$, Blanton et al. (2005)}. The first group contains the brightest members of the clusters, the third group contains the so-called dwarf population and the second one is formed by normal bright galaxies. Table 2 shows the median location, $r$-band absolute magnitude, velocity dispersion and local density\footnote{The local surface density ($\Sigma$) was computed with the 10 nearest neighbours to each galaxy belonging to the cluster.} of the different galaxy groups. In general, red galaxies are brighter than blue ones, and are also located closer to the cluster centre at higher local density regions. The two families of galaxies present different kinematics, in the sense that red galaxies show a smaller velocity dispersion than blue ones. This different kinematic between red and blue galaxies has also been seen in other studies, and have been interpreted as red and blue galaxies having different kind of orbits, being the orbits of  blue galaxies more anisotropic  than the red ones \citep[][]{adami98, biviano04}. Other authors interpret this difference in velocity dispersion as an evidence that ram pressure is not playing an important role in galaxy evolution in clusters. In contrast, tidal interactions should be the dominant mechanism \citep[][]{goto05}. All of these properties are independent of the sampled area.

It is also interesting that red dwarf galaxies are located at similar environments as the brightest red ones: close to the cluster centre in high local density regions \citep[see also ][]{hogg04}. But the red dwarf population shows a larger velocity dispersion than the brightest red galaxies. Biviano \& Katgert (2004) found that the brightest cluster members were not in equilibrium with the cluster potential. They are especial galaxies that could have formed close to the cluster centre or have fallen to this region due to dynamical friction. In contrast, dynamical friction is not so efficient in the dwarf population, so that the main presence of these galaxies in the central regions of the clusters should be due to their origin. The discussion about the properties and origin of the dwarf population will be given in another paper (S\'anchez-Janssen et al., in preparation).

%
\begin{table*}
\caption{Main properties of the different types of galaxies}             
\label{table:2}      
\centering          
\begin{tabular}{c c c c c c}
\hline\hline 
 Galaxies within $r/r_{200} < 5$ & $<r/r_{200}>$ & $<M_{r}>$ & $<\sigma>$ & $<log(\Sigma)>$ & $N_{gal}$ \\
\hline                    
$u-r<2.22$ & 1.85$\pm$0.02 & -19.65$\pm$0.01 & 1.04$\pm$0.01 & 0.46$\pm$0.08 & 4937 \\  
$u-r<2.22 \ \&  \ M_{r}<M_{r}^{*}-1$ & 1.79$\pm$0.11 & -21.54$\pm$0.02 & 1.08$\pm$0.09 & 0.40$\pm$0.25 & 94 \\
$u-r<2.22 \ \& \ M_{r}^{*}-1<M_{r}<M_{r}^{*}+1$ & 1.90$\pm$0.02 & -20.00$\pm$0.01 & 1.03$\pm$0.01 & 0.38$\pm$0.08 & 3126 \\
$u-r<2.22 \ \& \ M_{r}>M_{r}^{*}+1$ & 1.74$\pm$0.03 & -18.74$\pm$0.02 & 1.05$\pm$0.02 & 0.64$\pm$0.14 & 1717 \\
$u-r \ge 2.22$ & 1.03$\pm$0.01 & -20.16$\pm$0.01 & 0.90$\pm$0.01 & 0.80$\pm$0.05 & 5928 \\  
$u-r \ge 2.22 \ \& \ M_{r}<M_{r}^{*}-1$ & 0.95$\pm$0.05 & -21.62$\pm$0.02 & 0.78$\pm$0.03 & 0.89$\pm$0.14 & 537 \\
$u-r \ge 2.22 \ \& \ M_{r}^{*}-1<M_{r}<M_{r}^{*}+1$ & 1.10$\pm$0.02 & -20.20$\pm$0.01 & 0.91$\pm$0.01 & 0.75$\pm$0.06 & 4592 \\
$u-r \ge 2.22 \ \& \ M_{r}>M_{r}^{*}+1$ & 0.85$\pm$0.04 & -18.91$\pm$0.02 & 0.90$\pm$0.03 & 1.06$\pm$0.12 & 729 \\
\hline   \hline               
 Galaxies within $r/r_{200} < 2$ & $<r/r_{200}>$ & $<M_{r}>$ & $<\sigma>$ & $<log(\Sigma)>$ & $N_{gal}$ \\
\hline                    
$u-r<2.22$ & 0.97$\pm$0.01 & -19.61$\pm$0.02 & 1.08$\pm$0.02 & 0.80$\pm$0.07 & 2636 \\  
$u-r<2.22 \ \&  \ M_{r}<M_{r}^{*}-1$ & 1.15$\pm$0.07 & -21.56$\pm$0.03 & 1.18$\pm$0.13 & 0.62$\pm$0.24 & 54 \\
$u-r<2.22 \ \& \ M_{r}^{*}-1<M_{r}<M_{r}^{*}+1$ & 1.04$\pm$0.01 & -19.96$\pm$0.01 & 1.08$\pm$0.02 & 0.70$\pm$0.07 & 1648 \\
$u-r<2.22 \ \& \ M_{r}>M_{r}^{*}+1$ & 0.85$\pm$0.01 & -18.70$\pm$0.02 & 1.07$\pm$0.03 & 1.02$\pm$0.09 & 934 \\
$u-r \ge 2.22$ & 0.67$\pm$0.01 & -20.14$\pm$0.01 & 0.91$\pm$0.01 & 0.98$\pm$0.04 & 4244 \\  
$u-r \ge 2.22 \ \& \ M_{r}<M_{r}^{*}-1$ & 0.57$\pm$0.03 & -21.65$\pm$0.02 & 0.80$\pm$0.03 & 1.02$\pm$0.13 & 397 \\
$u-r \ge 2.22 \ \& \ M_{r}^{*}-1<M_{r}<M_{r}^{*}+1$ & 0.69$\pm$0.01 & -20.19$\pm$0.01 & 0.92$\pm$0.01 & 0.94$\pm$0.05 & 3239 \\
$u-r \ge 2.22 \ \& \ M_{r}>M_{r}^{*}+1$ & 0.61$\pm$0.02 & -18.92$\pm$0.02 & 0.89$\pm$0.03 & 1.19$\pm$0.10 & 608 \\
\hline                  

\end{tabular}
\end{table*}

   \begin{figure*}
   \centering
   \includegraphics{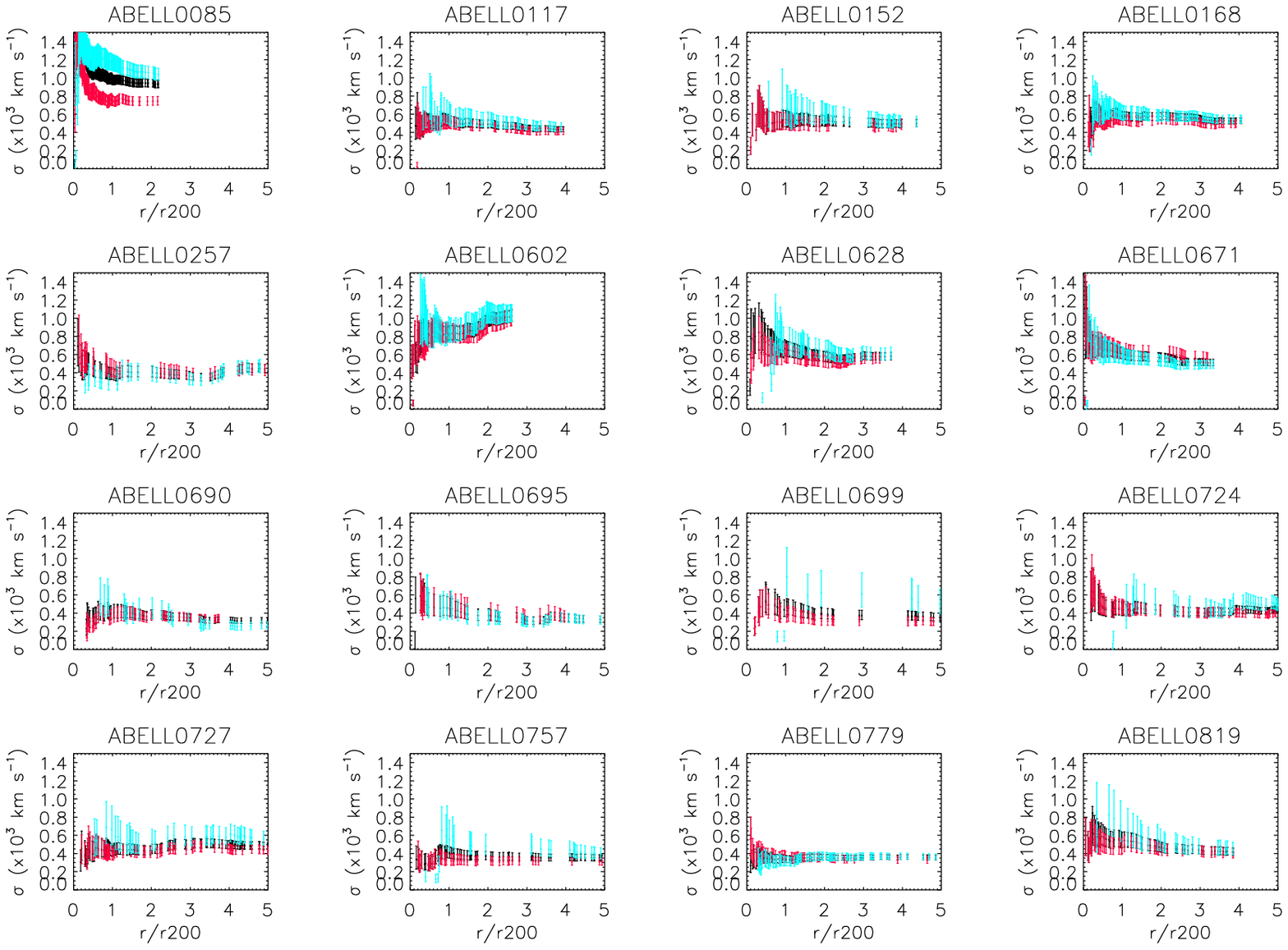}
      \caption{Velocity dispersion profiles of some clusters of our sample. The black symbols represent the velocity dispersion profile taking into account all types of galaxies. Blue and red symbols represent the velocity dispersion profiles corresponding to blue and red galaxies, respectively.
              }
   \end{figure*}
%

\section{Velocity Dispersion Profiles}

The adopted cluster velocity dispersion was calculated with the galaxies located within the $r_{200}$ radius of each cluster. But, how does $\sigma$ depend on the clustercentric distance in our sample?. This can be answered by studying the integrated velocity dispersion profiles (VDPs) of the clusters. These profiles also provide information about the dynamical properties of the galaxies. Thus, a system with galaxies predominantly in radial orbits produces an outwards declining VDP, while the opposite behaviour suggests instead that the galactic orbits are largely circular. In contrast, constant VDPs are characteristic of an isotropic distribution of velocities \citep[][]{solanes01}. Figure 6 shows the VDPs for some of the clusters in our sample. They show the velocity dispersion of the cluster at a given radius evaluated using all the galaxies within that radius, without any restriction in their luminosities. The errors showed in Fig. 6 were computed using the approximation given by Danese et al. (1980).

In order to classify the VDPs of our clusters, we computed the velocity dispersion ($\sigma_{i}, i=1,2,3,4,5$) of the galaxies in the clusters  located within five different radius: 0.4$\times$$r_{200}$, 0.6$\times$$r_{200}$, 2$\times$$r_{200}$, 3$\times$$r_{200}$ and 4$\times$$r_{200}$, respectively. We compared these values with $\sigma_{c}$,  given in Table 1. The resulting mean ratios $\sigma_{i}/\sigma_{c}$ were: 1.02$\pm$ 0.04, 1.01$\pm0.01$, 0.97$\pm0.01$, 0.94$\pm 0.02$ and 0.94$\pm$0.02, for $i=1,2,3,4,5$, respectively. These values indicate that within $r_{200}$ the VDPs of the total galaxy cluster population are consistent with being flat. The mean variation of the VDPs inside $r_{200}$ is only 2$\%$. The values of $\sigma_{i}/\sigma_{c}, i=3,4,5$ show that, outside $r_{200}$ the VDPs slowly decrease. The mean variation of the VDPs outside $r_{200}$ is $-6\%$. No differences in the ratios $\sigma_{i}/\sigma_{c}$ have been found when we have divided the galaxy sample between bright ($M_{r}<M_{r}^{*}+1$) and dwarf ($M_{r}>M_{r}^{*}+1$) galaxies. This flat behaviour of the VDPs inside $r_{200}$ suggests that galaxies in these areas have an isotropic distribution of velocities. In contrast, the decline with radius of VDPs outside $r_{200}$ points to radial orbits \citep[][]{solanes01}.

   \begin{figure}
   \centering
   \includegraphics[width=9cm,height=9cm]{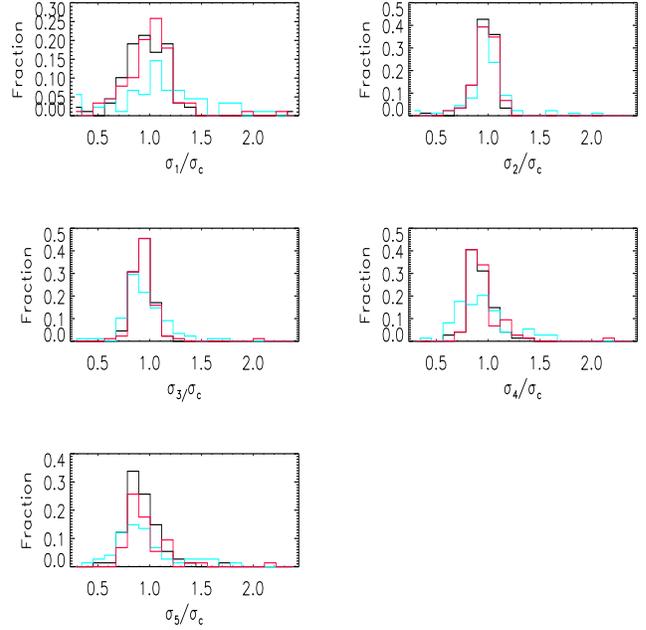}
      \caption{Histograms of the ratios $\sigma_{i}/\sigma_{c}, i=1,2,3,4,5$ for the galaxies in the clusters. The black full line represent all galaxies, the blue and red lines correspond to late- and early-type ones. See text for more details.}
   \end{figure}
%

Figure 6 also shows the VDPs of early- (red) and late-type (blue) galaxies. In most profiles the velocity dispersion of blue galaxies is larger than the corresponding one for early-type ones.  We have also analysed the shape of VDPs of blue and red galaxies as we did for the total sample. For red galaxies, we obtained that $\sigma_{i}/\sigma_{c,r}$ are 1.04$\pm$ 0.03, 1.03$\pm0.03$, 0.97$\pm0.02$, 0.96$\pm 0.02$ and 0.96$\pm$0.03, for $i=1, 2, 3, 4, 5$, respectively. The values of $\sigma_{i}/\sigma_{c,b}$ for the blue galaxies are: 1.15$\pm$ 0.07, 1.04$\pm0.03$, 0.95$\pm0.04$ and 0.93$\pm 0.04$ and 0.92$\pm$0.04, respectively. In those computations, $\sigma_{c,r}$ and $\sigma_{c,b}$ represent the velocity dispersion of the red and blue galaxies within a radius equal to $r_{200}$, respectively. Figure 7 show the distribution of $\sigma_{i}/\sigma_{c}, i=1,2,3,4,5$ for the blue, red and the total galaxy sample.

The VDPs have been studied in the literature by several authors. Most of them conclude that for large radii ($r>1$ Mpc) the VDPs are flat \citep[][]{girardi01, rines06, fadda96, muriel02}. This is consistent with the mild decrease that we have found in our clusters. The VDPs for red galaxies in our sample are almost flat outside $r_{200}$. This is not the case of the VDPs of blue galaxies which clearly decrease with radius outside $r_{200}$. In the inner regions ($r < r_{200}$) the VDPs of the total sample and those corresponding to the red galaxies are flat. In contrast, the VDPs of blue galaxies decrease with radius. Different authors show that  VDPs can decrease or increase with radius. den Hartog \& Katgert (1996) made a thorough study and found that the variations of the VDPs in the innermost regions of  clusters ($r < 0.5$ Mpc) are real and  not due to noise or bad centre election. We have re-computed $\sigma_{1}/\sigma_{c}$ and $\sigma_{2}/\sigma_{c}$ only for those clusters with X-ray centres, and our results did not significantly change. Thus, we can conclude that in our galaxy cluster sample only blue galaxies show increasing VDPs towards the centre of the cluster, while red galaxies show flat VDPs.

The previous findings can also be seen in Fig 8. We show the VDPs of the different galaxy classes for the normalised cluster, which was obtained by normalising the scales and velocities of each galaxy of the sample. Thus, the radial distance of each galaxy to the cluster centre was scaled by $r_{200}$ of the corresponding cluster, and the relative velocity of each galaxy cluster was normalised by the velocity dispersion of the cluster. Figure 8 shows the VDPs which correspond to the total, bright ($M_{r}<M_{r}^{*}+1$) and dwarf ($M_{r}>M_{r}^{*}+1$) galaxy samples. We have also distinguished between red and blue objects. The VDPs of the total galaxy sample indicate that blue galaxies have always  larger velocity dispersion than red ones. They also show always decreasing VDPs, while red ones have almost constant and slowly decrease VDPs inside and outside r$_{200}$, respectively. These features can also be seen in the VDPs of the bright galaxy sample. In contrast, red and blue dwarfs show decreasing VDPs inside r$_{200}$.

The shape of the VDPs can provide information about the dynamical state of the galaxies. Thus, clusters with galaxies predominantly in radial orbits produce an outwards declining VDP. This is the case of the blue galaxies of our sample, which is in agreement with previous findings \cite[][]{biviano04, adami98}. We have also obtained that the red dwarf galaxies inside r$_{200}$ has  an outwards declining VDP. This would  imply that this kind of galaxies may also be located in radial orbits. In contrast, constant VDPs imply an isotropic distribution of velocities \cite[][]{solanes01}. This is the case of the red bright galaxy population inside r$_{200}$.

   \begin{figure*}
   \centering
   \includegraphics[width=18cm]{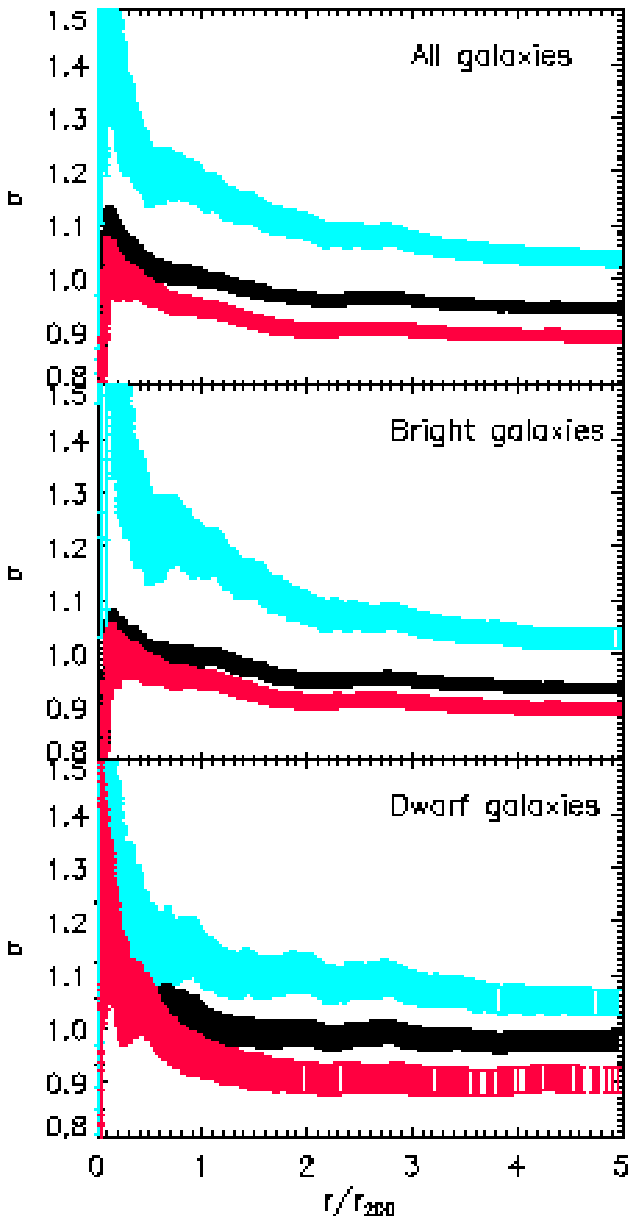}
      \caption{Velocity dispersion profiles of the galaxies of the normalised cluster. The total galaxy population is showed in the top panel. Bright galaxies ($M_{r}<M_{r}^{*}+1$) are in the middle panel, and the bottom panel shows the VDPs corresponding to the dwarf galaxy sample ($M_{r}>M_{r}^{*}+1$). The VDP of the total, blue and red galaxy samples are represented by black, blue and red colours, respectively (see text for more details).
              }
   \end{figure*}
%

\section{Fraction of blue galaxies}

Butcher \& Oemler (1984) observed that the fraction of blue galaxies ($f_{b}$) in clusters evolves with redshift, in the sense that galaxy clusters located at medium redshift have a larger $f_{b}$ than nearby ones. This has been usually interpreted as an evolutionary trend in clusters. But it is a matter of debate which is the role played by the environment in the change of the fraction of blue galaxies. We have computed $f_{b}$ in our sample of galaxy clusters, studying the variation with $z$ and the possible influence of the environment.

   \begin{figure*}[!htb]
   \centering
   \includegraphics[width=15cm]{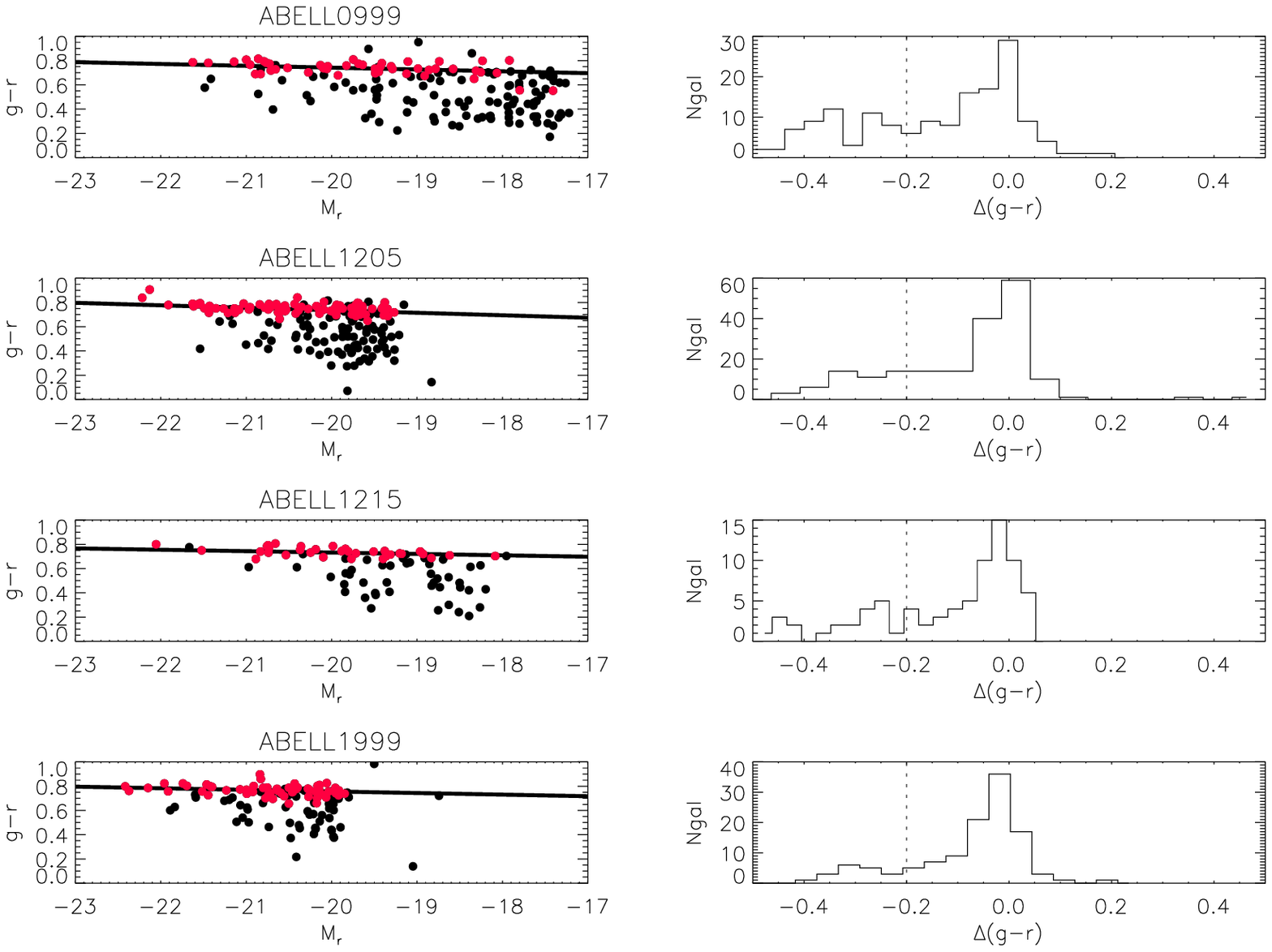}
      \caption{Colour-magnitude relation (left) and histograms (right) of marginalised colour distribution for four representative clusters at different redshifts of our cluster sample. The full line in left panels represent the fitted colour-magnitude relation. The vertical point lines in right panel represent the blue/red separation in the Butcher-Oemler effect. The red points are the galaxies with $u-r \ge 2.22$ and $C_{in}<0.4$ (see text for more details).
              }
   \end{figure*}
%

\subsection{Adopted aperture and limiting magnitude}

The original analysis of Butcher \& Oemler (1984) defined blue galaxies as those within a radius containing 30 per cent of the cluster population, being brighter than $M_{\rm{v}}=-20$ and bluer by 0.2 mag in $B-V$ than the colour-magnitude relation defined by the cluster early-type galaxies. It has been noticed by several authors that the fraction of blue galaxies strongly depends on the magnitude limit and the clustercentric distance used \citep[][]{ellingson01, goto03, depropris04, andreon06}. They observed that $f_{b}$ grows when the magnitude limit is fainter and the aperture is larger, reflecting the existence of a large fraction of blue faint galaxies in the outer regions of the clusters. De Propris et al. (2004) considered appropriate to measure $f_{b}$ in apertures based on cluster physical properties. They used $r_{200}$ as the aperture radius where they measured $f_{b}$ for their clusters. We have adopted also this radius in order to determine $f_{b}$ in our galaxy clusters. As it was previously commented, $f_{b}$ depends also on the adopted limiting magnitude of the galaxies in clusters. It should be noticed that as we move to higher redshifts we systematically lose faint galaxies (see Fig 2). Our clusters spread in a redshift range $0.02 < z < 0.1$, and only galaxies brighter than $M_{r}=-20.0$ ($\approx M_{r}^{*}+0.5$) can be observed at all redshifts. For this reason, we have adopted this absolute magnitude as the limiting magnitude for the computation of $f_{b}$. This ensures us to work with a complete galaxy sample at all redshifts. Other authors adopted fainter limiting magnitudes, e.g. $M^{*}+1.5$ \citep[][]{depropris04} or $M^{*}+3$ \citep[][]{margoniner00}. If there is a large number of blue galaxies at faint magnitudes, we expect that our values of $f_{b}$ will be smaller than those reported by the previous authors.

\subsection{Colour-magnitude diagrams}

We determined the $g-r$ versus $r$ colour-magnitude diagrams for all the clusters in our sample. The colour-magnitude relation was measured by a robust fitting routine by minimising the absolute deviation in $g-r$ colour, using only early-type galaxies located within an aperture of radius equal to $r_{200}$. The galaxy types were determined according to the $u-r$ colour and the light galaxy concentration parameter, $C_{in}$. These two criteria allow us to identify the most reliable sample of  E/S0 galaxies \citep[see ][]{shimasaku01, strateva01}. Thus, we considered early-type galaxies  those with $u-r \ge 2.22$ and $C_{in}<0.4$. Figure 9 shows the colour-magnitude diagrams of four representative galaxy clusters. The colour-magnitude relation fitted in each case is also overploted. Figure 9 (left column) also shows the histograms of the colour distribution, marginalised over the fitted colour-magnitude relation.

The average of the slopes of the colour-magnitude relations of the early-type galaxies of the clusters is -0.014$\pm$0.008. This slope is within the errors in agreement with the slope obtained by Gallazzi et al. (2006) for a large sample of galaxies using SDSS data. It is also in agreement with the average $B-R$ slope obtained by De Propris et al. (2004) for a sample of galaxy clusters from 2dFGRS.


\subsection{Calculation the blue fraction of galaxies}

As we explained before, the blue fraction of galaxies was computed using only those galaxies brighter than $M_{r} = -20$ and located within an aperture of radius $r_{200}$. In the present study we only used spectroscopically confirmed galaxy cluster members. This should not bias our results, especially due to our high completeness. Figure 10 presents $f_{b}$ as a function of redshift. The errors of $f_{b}$ were computed according to the prescription given by De Propris et al. (2004). We observe no evolution of $f_{b}$ with redshift, which means that our sample is ideal to study the effects of the environment on $f_{b}$.

   \begin{figure}
   \centering
   \includegraphics[width=9cm]{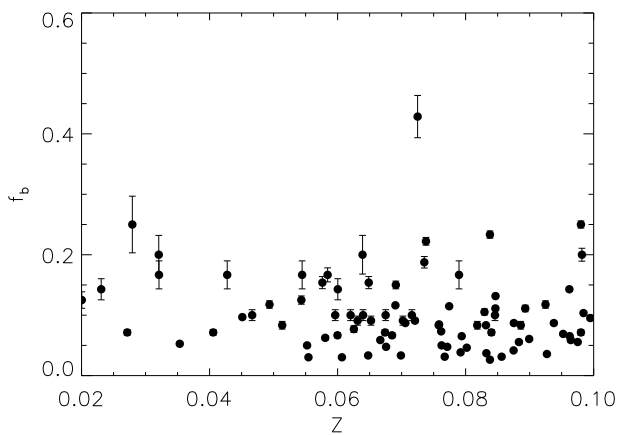}
      \caption{The fraction of blue galaxies ($f_{b}$) as a function of redshift of the galaxy clusters.
              }
   \end{figure}
%

We have considered three cluster properties (concentration, velocity dispersion and X-ray luminosity) of each cluster in order to analyse the dependence of $f_{b}$ on the environment. The concentration parameter was computed following the prescription of De Propris et al. (2004), i.e. $C = log(r_{60}/r_{20})$, where $r_{60}$ and $r_{20}$ are the radii containing 60 and 20 per cent of the cluster galaxies, respectively. The velocity dispersion of the clusters was taken from Table 1. The X-ray luminosities were obtained from the literature (Ebeling et al. 1998, B{\"o}hringer et al.2000, Ebeling et al. 2000 and Ledlow et al. 2003), being measured in the ROSAT band (0.1-2.4 keV). We only found X-ray data for 48 clusters of the sample.

Figure 11 shows the dependence of the fraction of blue galaxies on concentration, cluster velocity dispersion and X-ray luminosity. The non-parametric Spearman test returns that  $f_{b}$ has a low correlation with concentration and velocity dispersion. The fraction of blue galaxies correlates best with the velocity dispersion, but the significance of the correlation is 2.6$\sigma$. In contrast, the Spearman test shows correlation between $f_{b}$ and X-ray luminosity, being the significance of this correlation just 3$\sigma$. Notice that the points are distributed in the $L_{X}-f_{b}$ plane following a triangular shape. Clusters with large X-ray luminosity ($L_{X}(0.1-2.4keV) > 10^{45} erg s^{-1}$) show small fractions of blue galaxies (less than 10$\%$). Nevertheless, those clusters with small X-ray lumisosity show small and large fraction of blue galaxies. This correlation could indicate that there is a threshold over which cluster environment can affect the galaxy colours, and play a role in the galaxy evolution. This means that, according with our $L_{X}-\sigma$ relation, the evolution of galaxies could be driven by the cluster environment for those clusters with velocity dispersion larger than  $\sigma \approx 800$ km s$^{-1}$. Recently, \cite{popeso06} found a similar correlation between $L_{X}$ and $f_{b}$ for a larger cluster sample. The shape of the our $L_{X}-f_{b}$ correlation is similar to the correlation between cluster velocity dispersion and the fraction of [OII] emitters for clusters at low redshift reported by Poggianti et al. (2006). They found that clusters with $\sigma > 550$ km s$^{-1}$ have a constant low fraction (less than 30$\%$) of [OII] emiters. In contrast, those clusters with smaller $\sigma$ show large and small fractions of [OII] emiters.

We have recomputed $f_{b}$ taking into account those galaxies within an aperture of radius equal to $r_{200}$ and brighter than $M_{r}=-19.5$. We restricted the analysis only to those clusters with $z < 0.05$, because our sample is complete down to $M_{r}=-19.5$ in this redshift range. In this case the number of cluster decreases to 13. We have again studied the correlations of $f_{b}$ with  galaxy concentration, velocity dispersion and X-ray luminosity, obtaining similar correlations as with the full sample. 


   \begin{figure*}
   \centering
   \includegraphics[width=15cm]{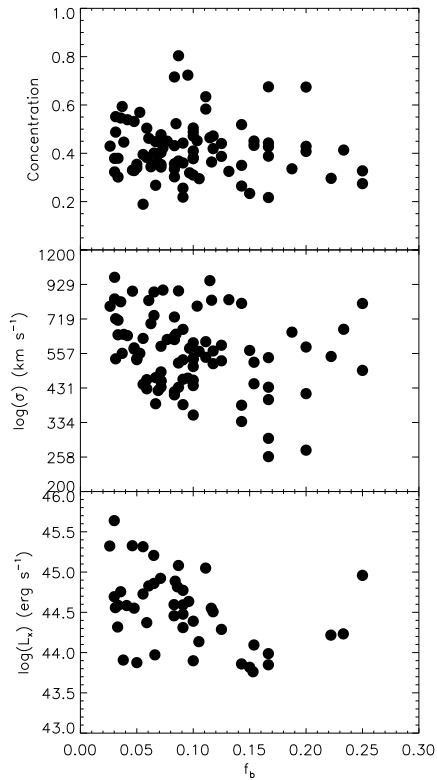}
      \caption{The fraction of blue galaxies ($f_{b}$) as a function of galaxy distribution (top), cluster velocity dispersion (middle), and X-ray luminosity (bottom) of the galaxy clusters.
              }
   \end{figure*}
%

\section{Discussion}

 From the study presented in this paper, most of the galaxies (62$\%$) located in the central regions of galaxy clusters ($r/r_{200}<2$) are early-type galaxies (see section 4). In contrast, the field population is dominated by late-type galaxies. In the literature it is also well established that the colour of galaxies in clusters and field is different, an indication of the low star formation activity found in cluster galaxies \citep[e.g. ][]{balogh98, lewis02, gomez03}. These differences in morphology and stellar content between field and cluster galaxies suggest different evolutionary processes. The facts that late-type galaxies show larger velocity dispersions and are located at larger distances from the cluster centre than early-type ones have been interpreted as late-type galaxies being recent arrivals to the cluster potential, forming a non-relaxed group of galaxies moving in more radial orbits than early-type ones \citep[e.g. ][]{stein97, adami98}. As late-type galaxies fall into the cluster potential and encounter denser environments, they evolve to early-type ones. The results presented in the present work are in agreement with previous findings. However, as pointed out by Goto (2005), this would imply that a large fraction of galaxies ($\approx40\%$ according to our sample) should be recent arrivals to the cluster, a possibility that seems unlikely. Goto (2005) concluded that the different observational properties between red and blue galaxies may indicate which is the main mechanism driving the evolution of galaxies in clusters. Gas stripping, mergers and interactions with other galaxies and with the cluster potential are the main mechanisms which are able to transform galaxies in clusters, making late-type galaxies lose their gas content, stop their star formation, circularise their orbits and transform their morphology from disk-like objects to spheroids. All of these mechanisms affect galaxies in clusters but, can we infer from the observational results which is the dominant one?.

It should be noted that the different mechanisms of galaxy evolution have very different time-scales. While gas stripping has a very short time-scale \citep[$\approx50$ Myr, ][]{quilis00}, the galaxy infall process can take $\approx1$ Gyr. The different mechanisms also have different underlying physics. Thus, ram pressure stripping is  proportional to the density of the intracluster medium (ICM) and to the square of the velocity of the galaxy. In contrast, dynamical interactions are more efficient when the relative velocity of galaxies is smaller \citep[][]{mamon92}. This means that gas stripping is stronger in the cluster centres and for galaxies with high velocities, while dynamical interactions should be more efficient for galaxies with smaller velocity dispersions. Numerical simulations have shown that most of the galaxies inside the virial radius have already been through the cluster core more than once \citep[][]{mamon04}. If gas stripping were the main mechanism driving galaxy evolution in clusters, according to the short time scale of this process,  only  few blue (late-type) galaxies should be observed in the central regions of clusters. Moreover, gas stripping is also stronger in galaxies with larger velocity dispersion which means that late-type galaxies should be more affected by this mechanism than early-type ones. Based on these considerations, Goto (2005) concluded that gas stripping is not the main responsible mechanism driving the evolution of galaxies in clusters. Instead, galaxies in clusters evolve mainly by dynamical interactions. We can add to Goto's discussion that if gas stripping were the main galactic evolution mechanism in clusters, then the fraction of blue galaxies should depend on the cluster mass as the temperature and density of the gas increases with the cluster mass. According to our results, this is true for those clusters with large X-ray luminosities. In contrast, the cluster environment is not so important in driving the evolution of galaxies in low mass clusters. Thus, gas stripping may not be the main responsible mechanism transforming late-type to early-type galaxies in low mass clusters, but could be important in the most massive ones. This does not mean that gas stripping is absent in the evolution of galaxies in clusters; some clear examples of gas stripping have been observed in galaxies in Virgo \citep[][]{kenney04}.

Dynamical interactions include both interactions with the cluster potential and with other galaxies. These effects can trigger  temporary star formation  in cluster galaxies \citep[][]{fujita98}, which can be analysed by studying their colour distribution. These interactions can also disrupt stars from galaxies, forming at the beginning long tidal tails that subsequently will be diluted and will form the diffuse light observed in some nearby clusters like Virgo \citep[see ][ and references therein]{aguerri05b}.  These effects will be more important in those galaxies with smaller relative velocities. Fujita (1998) conclude that if the tidal effects enhance the SFR in the galaxies, then the bluest galaxies should be located close to the cluster centre (within $\approx 300$ kpc), whereas they should be in the outer parts of the cluster if the SFR is induced by galaxy-galaxy encounters. We have investigated the fraction of blue galaxies in our clusters located within 300 kpc from the centre of the cluster. The sample has been divided in bright and dwarf galaxies ($M_{r}<M^{*}_{r}+1$ and $M_{r}>M^{*}_{r}+1$, respectively). We have obtained that 40$\%$ of the blue bright galaxies and 30$\%$ of the blue dwarf ones are located at smaller distance than 300 kpc from the cluster center. This means that tidal interactions with the cluster potential are not the responsible mechanism for the formation of most of the blue galaxies in our clusters. The lack of blue galaxies in the central regions of clusters has been observed also in nearby clusters like Coma \citep[][]{aguerri04} as well as in other distant clusters \citep[][]{rakos97, abraham96, balogh97}. 

These evidences indicate that the evolution of galaxies in clusters could be driven by the cluster environment in the most massive ones, but galaxies in low mass clusters could mainly evolve due to the local environment.


\section{Conclusions}

In the present paper we have analysed the main properties of the galaxies of one of the largest (10865 galaxies) and homogeneous sample presented in the literature. The galaxies have been grouped in two families according to their $u-r$ colour. Those galaxies with $u-r \ge 2.22$ formed the red (early-type) family, and those with $u-r<2.22$ the blue (late-type) one.  We have derived the position, velocity dispersion, and VDPs of both families of galaxies, obtaining: 

\begin{itemize}
\item Within 2$\times$$r_{200}$, 62$\%$ and 38$\%$ of the galaxies turned to be red and blue, respectively.
\item The median positions and velocity dispersions are smaller for red galaxies than for blue ones.
\item Bright ($M_{r}<M_{r}^{*}-1$) and dwarf ($M_{r}>M_{r}^{*}+1$) red galaxies are located at smaller distances than the blue ones, sharing  the same cluster environment.
\item The brightest cluster members ($M_{r}<-21.0$) show smaller velocity dispersions than the remaining.
\item The VDPs of the total galaxy cluster population are constant with radius in the central regions of the clusters ($r<r_{200}$), while slowly decrease in the outermost regions ($r \ge r_{200}$). The red galaxy population have also flat VDPs in the central regions ($r<r_{200}$). In contrast, the VDPs of blue galaxies grow towards the cluster centre. In the outer regions ($r>r_{200}$), the VDPs of red galaxies decline smoothly with radius, while for blue ones the decrement is faster. This indicates that the galaxies in the outermost regions of the clusters are dominated by the blue population, and have more radial and anisotropic orbits than galaxies in the inner regions dominated by the red population.
\item The fraction of blue galaxies in our cluster sample does not correlate with cluster global properties, such as the concentration of the galaxy distribution and cluster velocity dispersion. However, we found a correlation between the X-ray luminosity and the fraction of blue galaxies. Those clusters with $L_{X}(0.1-2.4 keV) > 10^{45}$ erg s$^{-1}$ have a low fraction of blue galaxies (less than 10$\%$). In contrast, clusters with low of X-ray luminosity show large and small fractions of blue galaxies. This could indicate that the star formation in cluster galaxies may be regulated by global cluster properties for clusters with $L_{X}(0.1-2.4 keV) > 10^{45}$ erg s$^{-1}$, i.e. those clusters with $\sigma_{c}> 800$ km s$^{-1}$.
\end{itemize}

All these results are in agreement with previous findings from other cluster samples, indicating that red and blue galaxies have different evolution in galaxy clusters. We have discussed these results according to the different galaxy transformation mechanisms presented in galaxy clusters, concluding the local environment plays a key role in galaxy evolution in low mass clusters, while the evolution of galaxies in massive clusters could be driven by the global cluster environment.

\begin{acknowledgements}
We wish to thank to the anonymous referee for useful coments which have improved this manuscript. We also acknowledge  financial support by the Spanish Ministerio de Ciencia y Tecnolog\'{\i}a grants AYA2004-08260.  We would like also to thank T. Beers for providing us with copy of his code ROSTAT, and K. M. Ashman and S. Zepf for making their KMM code available to us. Funding for the SDSS and SDSS-II has been provided by the Alfred P. Sloan Foundation, the Participating Institutions, the National Science Foundation, the U.S. Department of Energy, the National Aeronautics and Space Administration, the Japanese Monbukagakusho, the Max Planck Society, and the Higher Education Funding Council for England. The SDSS Web Site is http://www.sdss.org/. The SDSS is managed by the Astrophysical Research Consortium for the Participating Institutions. The Participating Institutions are the American Museum of Natural History, Astrophysical Institute Potsdam, University of Basel, Cambridge University, Case Western Reserve University, University of Chicago, Drexel University, Fermilab, the Institute for Advanced Study, the Japan Participation Group, Johns Hopkins University, the Joint Institute for Nuclear Astrophysics, the Kavli Institute for Particle Astrophysics and Cosmology, the Korean Scientist Group, the Chinese Academy of Sciences (LAMOST), Los Alamos National Laboratory, the Max-Planck-Institute for Astronomy (MPIA), the Max-Planck-Institute for Astrophysics (MPA), New Mexico State University, Ohio State University, University of Pittsburgh, University of Portsmouth, Princeton University, the United States Naval Observatory, and the University of Washington. This research has made use of the NASA/IPAC Extragalactic Database (NED) which is operated by the Jet Propulsion Laboratory, California Institute of Technology, under contract with the National Aeronautics and Space Administration.

\end{acknowledgements}

\begin{longtable}{cccccccc}
\caption{Cluster characteristics} \\
\hline\hline
	Name& $\alpha$ (J2000)& $\delta$ (J2000)& v$_{c}$ & $\sigma_{c}$ &r$_{200}$ & N$_{gal}$ & C \\
      & (degrees)& (degrees)& (km s$^{-1}$) & (km s$^{-1}$) & (Mpc) &  &   \\
\hline
\endfirsthead
\caption{continued.}\\
\hline\hline
	Name& $\alpha$ (J2000)& $\delta$ (J2000)& v$_{c}$ & $\sigma_{c}$ & r$_{200}$ & N$_{gal}$ & C \\
      & (degrees)& (degrees)& (km s$^{-1}$) & (km s$^{-1}$) & (Mpc) &  &   \\
\hline
\hline
\endhead
\hline
\endfoot
Abell0085&      10.4571&     -9.30694&   16633$^{+      40}_{-      29}$ &     979$^{+      42}_{-      39}$ &      2.10&     273&     0.93\\
Abell0117&      14.0100&     -10.0022&   16568$^{+      31}_{-      42}$ &     531$^{+      31}_{-      27}$ &      1.19&      60&     0.88\\
Abell0152&      17.5229&      13.9804&   17888$^{+      67}_{-      34}$ &     538$^{+      43}_{-      38}$ &      1.12&      27&     0.85\\
Abell0168&      18.7429&     0.365833&   13534$^{+      23}_{-      14}$ &     578$^{+      31}_{-      28}$ &      1.19&     106&     0.88\\
Abell0257&      27.3396&      14.0372&   21060$^{+      47}_{-      21}$ &     381$^{+      52}_{-      44}$ &     0.81&      26&     0.94\\
Abell0602&      118.341&      29.3717&   18202$^{+      34}_{-      51}$ &     834$^{+      68}_{-      61}$ &      1.87&      78&     0.74\\
Abell0628&      122.543&      35.2958&   25139$^{+      24}_{-      89}$ &     666$^{+      43}_{-      38}$ &      1.47&      43&     0.83\\
Abell0671&      127.121&      30.4169&   14599$^{+      19}_{-      33}$ &     610$^{+      37}_{-      33}$ &      1.42&      72&     0.89\\
Abell0690&      129.815&      28.9033&   23689$^{+      44}_{-      23}$ &     395$^{+      28}_{-      24}$ &     0.85&      22&     0.95\\
Abell0695&      130.309&      32.4174&   20251$^{+      46}_{-      37}$ &     456$^{+      39}_{-      32}$ &      1.04&      16&     0.86\\
Abell0699&      131.236&      27.7508&   25375$^{+      35}_{-      49}$ &     438$^{+      45}_{-      37}$ &     0.91&      19&     0.73\\
Abell0724&      134.600&      38.5137&   28134$^{+      25}_{-      55}$ &     433$^{+      36}_{-      32}$ &     1.00&      29&     0.94\\
Abell0727&      134.976&      39.4389&   28571$^{+      54}_{-      14}$ &     423$^{+      33}_{-      29}$ &     0.96&      33&     0.96\\
Abell0757&      138.277&      47.7036&   15402$^{+      27}_{-      36}$ &     409$^{+      35}_{-      30}$ &     0.84&      30&     0.85\\
Abell0779&      139.962&      33.7714&    6921$^{+      13}_{-      33}$ &     336$^{+      24}_{-      21}$ &     0.79&      57&     0.75\\
Abell0819&      143.076&      9.68861&   22872$^{+      39}_{-      50}$ &     536$^{+      44}_{-      37}$ &      1.19&      31&     0.94\\
Abell0883&      147.822&      5.48799&   21750$^{+     109}_{-      30}$ &     523$^{+      77}_{-      58}$ &      1.17&      18&     0.91\\
Abell0971&      154.997&      40.9925&   27809$^{+      54}_{-      43}$ &     816$^{+      72}_{-      61}$ &      1.88&      40&     0.87\\
Abell0999&      155.842&      12.8466&    9618$^{+      10}_{-      54}$ &     271$^{+      19}_{-      17}$ &     0.60&      25&     0.90\\
Abell1003&      156.235&      47.8442&   18762$^{+      44}_{-      84}$ &     617$^{+      39}_{-      34}$ &      1.37&      29&     0.94\\
Abell1016&      156.762&      10.9780&    9629$^{+      34}_{-       9}$ &     259$^{+      20}_{-      17}$ &     0.60&      25&     0.91\\
Abell1024&      157.096&      3.76341&   22067$^{+      40}_{-      20}$ &     532$^{+      40}_{-      34}$ &      1.26&      35&     0.90\\
Abell1032&      157.547&      4.03417&   20008$^{+      26}_{-      24}$ &     355$^{+      38}_{-      32}$ &     0.77&      25&     0.89\\
Abell1035&      158.092&      40.1817&   20270$^{+      36}_{-      34}$ &     575$^{+      52}_{-      45}$ &      1.34&      49&     0.97\\
Abell1066&      159.911&      5.17444&   20708$^{+       4}_{-      81}$ &     826$^{+      49}_{-      44}$ &      1.71&      95&     0.92\\
Abell1142&      165.229&      10.5477&   10601$^{+      30}_{-      25}$ &     557$^{+      43}_{-      38}$ &      1.33&      59&     0.88\\
Abell1149&      165.769&      7.57833&   21479$^{+       2}_{-      64}$ &     352$^{+      37}_{-      31}$ &     0.85&      26&     0.94\\
Abell1169&      166.967&      43.9506&   17532$^{+      24}_{-      35}$ &     433$^{+      37}_{-      32}$ &     0.91&      35&     0.94\\
Abell1173&      167.328&      41.5624&   22789$^{+      23}_{-      69}$ &     611$^{+      46}_{-      41}$ &      1.33&      35&     0.95\\
Abell1189&      167.775&      1.09899&   28860$^{+     109}_{-      60}$ &     807$^{+     104}_{-      88}$ &      1.76&      41&     0.94\\
Abell1190&      167.902&      40.8417&   22610$^{+      17}_{-      39}$ &     706$^{+      33}_{-      30}$ &      1.50&      77&     0.92\\
Abell1205&      168.328&      2.53867&   22852$^{+      17}_{-      53}$ &     890$^{+      59}_{-      53}$ &      2.04&      83&     0.84\\
Abell1215&      170.100&      4.34280&   14747$^{+      25}_{-      10}$ &     214$^{+      46}_{-      39}$ &     0.48&      17&     0.89\\
Abell1238&      170.711&      1.09389&   22140$^{+       9}_{-      54}$ &     544$^{+      48}_{-      41}$ &      1.30&      60&     0.91\\
Abell1270&      172.366&      54.0428&   20728$^{+      49}_{-      36}$ &     569$^{+      45}_{-      39}$ &      1.31&      43&     0.99\\
Abell1291&      173.092&      55.9783&   17144$^{+      43}_{-      60}$ &     720$^{+      39}_{-      36}$ &      1.58&      45&     0.97\\
Abell1318&      173.883&      55.0767&   17185$^{+      54}_{-      22}$ &     360$^{+      28}_{-      25}$ &     0.81&      22&     0.96\\
Abell1346&      175.304&      5.74613&   29523$^{+      41}_{-      25}$ &     790$^{+      70}_{-      62}$ &      1.69&      66&     0.85\\
Abell1377&      176.883&      55.7597&   15378$^{+      35}_{-      33}$ &     671$^{+      42}_{-      37}$ &      1.47&      69&     0.96\\
Abell1383&      176.973&      54.7089&   17855$^{+      38}_{-      35}$ &     456$^{+      26}_{-      23}$ &      1.02&      35&     0.95\\
Abell1385&      177.017&      11.5864&   25337$^{+      49}_{-      44}$ &     609$^{+      52}_{-      45}$ &      1.34&      22&     0.86\\
Abell1390&      177.378&      12.3034&   25101$^{+      48}_{-      29}$ &     483$^{+      47}_{-      41}$ &      1.06&      27&     0.86\\
Abell1424&      179.361&      5.12000&   22736$^{+      33}_{-      39}$ &     617$^{+      49}_{-      43}$ &      1.46&      63&     0.95\\
Abell1436&      180.095&      56.2314&   19432$^{+      39}_{-      77}$ &     712$^{+      50}_{-      44}$ &      1.51&      66&     0.96\\
Abell1452&      180.802&      51.6642&   18609$^{+      56}_{-      58}$ &     533$^{+      35}_{-      30}$ &      1.30&      18&     0.91\\
Abell1459&      181.108&      1.88281&    6010$^{+      37}_{-      21}$ &     527$^{+      58}_{-      48}$ &      1.18&      65&     0.95\\
Abell1507&      183.766&      59.8947&   18009$^{+       9}_{-      70}$ &     379$^{+      41}_{-      36}$ &     0.85&      23&     0.91\\
Abell1516&      184.729&      5.24731&   23019$^{+      24}_{-      84}$ &     720$^{+      48}_{-      43}$ &      1.49&      60&     0.90\\
Abell1552&      187.392&      11.7733&   26495$^{+      64}_{-      42}$ &     442$^{+      37}_{-      32}$ &     0.93&      20&     0.88\\
Abell1564&      188.720&      1.78056&   23763$^{+      30}_{-      64}$ &     641$^{+      72}_{-      62}$ &      1.51&      46&     0.91\\
Abell1616&      191.817&      55.0006&   24882$^{+      65}_{-      93}$ &     565$^{+      52}_{-      45}$ &      1.32&      29&     0.87\\
Abell1620&      192.510&     -1.53764&   25400$^{+      64}_{-      38}$ &     829$^{+      49}_{-      43}$ &      1.76&      58&     0.88\\
Abell1630&      192.942&      4.59694&   19458$^{+      36}_{-      33}$ &     444$^{+      45}_{-      37}$ &     0.98&      30&     0.90\\
Abell1650&      194.672&     -1.76417&   25138$^{+      86}_{-      18}$ &     790$^{+      53}_{-      47}$ &      1.61&      63&     0.85\\
Abell1663&      195.717&     -2.51782&   24953$^{+      60}_{-      20}$ &     729$^{+      44}_{-      40}$ &      1.54&      72&     0.88\\
Abell1692&      198.060&    -0.976000&   25395$^{+      51}_{-      47}$ &     607$^{+      49}_{-      43}$ &      1.32&      40&     0.89\\
Abell1728&      200.876&      11.2960&   26977$^{+      95}_{-      22}$ &     824$^{+      72}_{-      62}$ &      1.88&      50&     0.70\\
Abell1750&      202.709&     -1.86389&   26259$^{+      19}_{-      14}$ &     518$^{+      37}_{-      33}$ &      1.15&      35&     0.95\\
Abell1767&      204.024&      59.2042&   21174$^{+      39}_{-      32}$ &     885$^{+      44}_{-      40}$ &      2.05&     109&     0.94\\
Abell1773&      205.533&      2.24805&   23544$^{+      31}_{-      48}$ &     481$^{+      36}_{-      31}$ &      1.09&      32&     0.87\\
Abell1780&      206.149&      2.86750&   23285$^{+      35}_{-      22}$ &     624$^{+      59}_{-      53}$ &      1.45&      53&     0.90\\
Abell1783&      205.848&      55.6261&   20550$^{+      48}_{-      11}$ &     383$^{+      36}_{-      32}$ &     0.94&      33&     0.94\\
Abell1809&      208.245&      5.16139&   23815$^{+      39}_{-      31}$ &     737$^{+      53}_{-      47}$ &      1.68&      89&     0.82\\
Abell1885&      213.431&      43.6634&   26793$^{+      33}_{-      50}$ &     541$^{+      72}_{-      56}$ &      1.23&      22&     0.92\\
Abell1999&      223.522&      54.2682&   29841$^{+      11}_{-      57}$ &     463$^{+      62}_{-      54}$ &      1.07&      24&     0.92\\
Abell2018&      225.266&      47.2831&   26246$^{+      93}_{-       5}$ &     635$^{+      43}_{-      37}$ &      1.45&      39&     0.88\\
Abell2023&      227.496&      2.98910&   27743$^{+      48}_{-      18}$ &     516$^{+      88}_{-      73}$ &      1.12&      23&     0.86\\
Abell2026&      227.106&    -0.267500&   27188$^{+      62}_{-      37}$ &     747$^{+      57}_{-      49}$ &      1.50&      43&     0.91\\
Abell2030&      227.844&   -0.0857717&   27399$^{+      38}_{-      27}$ &     495$^{+      51}_{-      45}$ &      1.10&      38&     0.91\\
Abell2061&      230.317&      30.6122&   23646$^{+      24}_{-      19}$ &     622$^{+      35}_{-      32}$ &      1.43&      98&     0.83\\
Abell2067&      230.780&      30.8703&   23039$^{+      33}_{-      34}$ &     917$^{+      50}_{-      46}$ &      2.19&     118&     0.85\\
Abell2092&      233.348&      31.1475&   20000$^{+      42}_{-      17}$ &     458$^{+      35}_{-      30}$ &     0.93&      41&     0.86\\
Abell2110&      234.953&      30.7173&   29250$^{+      94}_{-      28}$ &     622$^{+      59}_{-      49}$ &      1.25&      21&     0.83\\
Abell2122&      236.259&      36.1161&   19793$^{+      28}_{-      42}$ &     826$^{+      52}_{-      47}$ &      1.80&      91&     0.92\\
Abell2124&      236.263&      36.1172&   19783$^{+      41}_{-      32}$ &     826$^{+      52}_{-      47}$ &      1.78&      90&     0.92\\
Abell2145&      240.094&      33.2306&   26583$^{+      80}_{-      23}$ &     632$^{+      69}_{-      56}$ &      1.40&      24&     0.87\\
Abell2149&      240.350&      53.9061&   19564$^{+      63}_{-      44}$ &     459$^{+      31}_{-      27}$ &      1.01&      20&     0.85\\
Abell2169&      243.422&      49.1261&   17286$^{+      30}_{-      50}$ &     521$^{+      38}_{-      33}$ &      1.24&      40&     0.82\\
Abell2175&      245.132&      29.8953&   28876$^{+      61}_{-      64}$ &     878$^{+      66}_{-      57}$ &      1.76&      58&     0.87\\
Abell2199&      247.154&      39.5244&    9118$^{+      15}_{-      30}$ &     747$^{+      20}_{-      19}$ &      1.77&     269&     0.92\\
Abell2241&      254.928&      32.6161&   29403$^{+      70}_{-      68}$ &     806$^{+      73}_{-      62}$ &      1.61&      37&     0.90\\
Abell2244&      255.663&      34.0411&   28927$^{+      50}_{-      57}$ &     428$^{+      59}_{-      49}$ &     0.99&      23&     0.94\\
Abell2245&      255.640&      33.5056&   25686$^{+      40}_{-      49}$ &     535$^{+      44}_{-      39}$ &      1.28&      39&     0.95\\
Abell2255&      258.222&      64.0653&   24052$^{+      19}_{-      39}$ &     883$^{+      38}_{-      35}$ &      1.86&     184&     0.91\\
Abell2428&      334.065&     -9.34139&   25207$^{+      15}_{-      68}$ &     433$^{+      44}_{-      37}$ &     0.95&      26&     0.87\\
Abell2670&      358.556&     -10.4133&   22755$^{+      29}_{-      20}$ &     642$^{+      32}_{-      29}$ &      1.46&     137&     0.94\\
MACSJ0810.3+4216&      122.600&      42.2733&   19193$^{+      44}_{-      25}$ &     505$^{+      42}_{-      35}$ &      1.23&      32&     0.84\\
MACSJ1440.0+3707&      220.011&      37.0839&   29402$^{+      88}_{-      46}$ &     587$^{+      62}_{-      49}$ &      1.31&      18&     0.91\\
NSCJ152902+524945&      232.309&      52.8433&   22063$^{+     101}_{-       7}$ &     652$^{+      51}_{-      43}$ &      1.4&      45&     0.89\\
NSCJ161123+365846&      242.854&      36.9700&   20221$^{+      39}_{-      25}$ &     485$^{+      45}_{-      39}$ &      1.17&      30&     0.86\\
RBS1385&      215.969&      40.2619&   24544$^{+      40}_{-      78}$ &     419$^{+      44}_{-      36}$ &     0.84&      16&     0.91\\
RXCJ0137.2-0912&      24.3137&     -9.20277&   12169$^{+      27}_{-      30}$ &     453$^{+      35}_{-      30}$ &     0.93&      49&     0.92\\
RXCJ0828.6+3025&      127.162&      30.4280&   14630$^{+      50}_{-      25}$ &     628$^{+      37}_{-      33}$ &      1.45&      76&     0.89\\
RXCJ0953.6+0142&      148.393&      1.70550&   29450$^{+      24}_{-      63}$ &     584$^{+      70}_{-      59}$ &      1.30&      22&     0.96\\
RXCJ1115.5+5426&      168.887&      54.4350&   20965$^{+      35}_{-      66}$ &     639$^{+      43}_{-      38}$ &      1.35&      50&     0.94\\
RXCJ1121.7+0249&      170.428&      2.81840&   14807$^{+      25}_{-      14}$ &     567$^{+      46}_{-      41}$ &      1.40&      73&     0.85\\
RXCJ1351.7+4622&      207.940&      46.3668&   18937$^{+      32}_{-      40}$ &     531$^{+      30}_{-      27}$ &      1.16&      40&     0.94\\
RXCJ1424.8+0240&      216.159&      2.75677&   16337$^{+      44}_{-      57}$ &     539$^{+      44}_{-      36}$ &      1.19&      22&     0.91\\
RXJ1017.7-0002&      154.452&   -0.0595327&   19169$^{+      44}_{-      28}$ &     413$^{+      47}_{-      39}$ &     0.90&      16&     0.91\\
RXJ1022.1+3830&      155.583&      38.5308&   16300$^{+      37}_{-      38}$ &     591$^{+      38}_{-      33}$ &      1.35&      51&     0.82\\
RXJ1053.7+5450&      163.449&      54.8500&   21623$^{+      23}_{-      49}$ &     665$^{+      51}_{-      45}$ &      1.52&      46&     0.86\\
WBL238&      146.732&      54.4183&   13995$^{+      51}_{-      17}$ &     602$^{+      34}_{-      30}$ &      1.30&      44&     0.87\\
WBL518&      220.179&      3.45305&    8141$^{+      19}_{-      28}$ &     454$^{+      22}_{-      20}$ &      1.03&     103&     0.85\\
ZwCl0027.0-0036&      7.31721&    -0.183598&   17994$^{+      18}_{-      44}$ &     465$^{+      42}_{-      36}$ &     0.99&      36&     0.85\\
ZwCl0743.5+3110&      116.655&      31.0136&   17419$^{+      49}_{-      89}$ &     694$^{+      51}_{-      45}$ &      1.40&      29&     0.82\\
ZwCl1207.5+0542&      182.578&      5.38500&   23137$^{+      43}_{-      36}$ &     580$^{+      46}_{-      40}$ &      1.20&      40&     0.91\\
ZwCl1215.1+0400&      184.422&      3.66040&   23229$^{+      22}_{-      40}$ &     955$^{+      43}_{-      39}$ &      2.17&     130&     0.90\\
ZwCl1316.4-0044&      199.816&    -0.907816&   24972$^{+      56}_{-      31}$ &     557$^{+      22}_{-      20}$ &      1.16&      38&     0.87\\
ZwCl1730.4+5829&      261.856&      58.4749&    8379$^{+      26}_{-      36}$ &     491$^{+      24}_{-      22}$ &      1.02&      33&     0.86\\
\hline
\end{longtable}


\begin{thebibliography}{plainnat}

\bibitem[Abell et al.(1989)]{abell89} Abell, G.~O., Corwin,
 H.~G., Jr., \& Olowin, R.~P.\ 1989, \apjs, 70, 1


\bibitem[Abraham et al.  1996]{abraham96} Abraham, R.~G., et al.\ 
1996, \apj, 471, 694

\bibitem[Adami et al.  1998]{adami98} Adami, C., Biviano, A., 
\& Mazure, A.\ 1998, \aap, 331, 439

\bibitem[Adelman-McCarthy et al. 2006]{adelman06} 
Adelman-McCarthy, J.~K., et al.\ 2006, \apjs, 162, 38 


\bibitem[Aguerri et al. 2006]{aguerri06} Aguerri, J.~A.~L., Castro-Rodr\'{\i}guez, N., Napolitano, N., Arnaboldi, M. \& Gerhard, O., 2006, astro-ph/0607263

\bibitem[Aguerri et al. 2005a]{aguerri05a} Aguerri, J.~A.~L., 
Iglesias-P{\'a}ramo, J., V{\'{\i}}lchez, J.~M., Mu{\~n}oz-Tu{\~n}{\'o}n, 
C., \& S{\'a}nchez-Janssen, R.\ 2005a, \aj, 130, 475 

\bibitem[Aguerri et al.  2005b]{aguerri05b} Aguerri, J.~A.~L., 
Gerhard, O.~E., Arnaboldi, M., Napolitano, N.~R., Castro-Rodriguez, N., \& 
Freeman, K.~C.\ 2005b, \aj, 129, 2585



\bibitem[Aguerri et al.  2004]{aguerri04} Aguerri, J.~A.~L., 
Iglesias-Paramo, J., Vilchez, J.~M., \& Mu{\~n}oz-Tu{\~n}{\'o}n, C.\ 2004, 
\aj, 127, 1344

\bibitem[Akritas \& Bershady  1996]{akritas96} Akritas, M.~G., \& 
Bershady, M.~A.\ 1996, \apj, 470, 706 

\bibitem[Andreon et al.  2006]{andreon06} Andreon, S., Quintana, 
H., Tajer, M., Galaz, G., \& Surdej, J.\ 2006, \mnras, 365, 915 

\bibitem[Andreon \& Ettori  1999]{andreon99} Andreon, S., \& 
Ettori, S.\ 1999, \apj, 516, 647 

\bibitem[Andreon et al.  1996]{andreon96} Andreon, S., Davoust, 
E., Michard, R., Nieto, J.-L., \& Poulain, P.\ 1996, \aaps, 116, 429

\bibitem[Arnaboldi et al. 2004 ]{arnaboldi04} Arnaboldi, M., 
Gerhard, O., Aguerri, J.~A.~L., Freeman, K.~C., Napolitano, N.~R., Okamura, 
S., \& Yasuda, N.\ 2004, \apjl, 614, L33 

\bibitem[Arnaboldi et al.  2002]{arnaboldi02} Arnaboldi, M., et 
al.\ 2002, \aj, 123, 760 


\bibitem[Ashman et al.  1994]{ashman94} Ashman, K.~M., Bird, 
C.~M., \& Zepf, S.~E.\ 1994, \aj, 108, 2348


\bibitem[Balogh et al.  1998]{balogh98} Balogh, M.~L., Schade, 
D., Morris, S.~L., Yee, H.~K.~C., Carlberg, R.~G., \& Ellingson, E.\ 1998, 
\apjl, 504, L75

\bibitem[Balogh et al.  1997]{balogh97} Balogh, M.~L., Morris, 
S.~L., Yee, H.~K.~C., Carlberg, R.~G., \& Ellingson, E.\ 1997, \apjl, 488, 
L75

\bibitem[Beers et al.  1990]{beers90} Beers, T.~C., Flynn, K., 
\& Gebhardt, K.\ 1990, \aj, 100, 32 

\bibitem[Bekki et al. 2002]{bekki02} Bekki, K., Couch, W.~J., 
\& Shioya, Y.\ 2002, \apj, 577, 651 


\bibitem[Biviano \& Katgert  2004]{biviano04} Biviano, A., \& 
Katgert, P.\ 2004, \aap, 424, 779 

\bibitem[Biviano \& Girardi 2003]{biviano03} Biviano, A., \& 
Girardi, M.\ 2003, \apj, 585, 205

\bibitem[Biviano et al.  1992]{biviano92} Biviano, A., Girardi, 
M., Giuricin, G., Mardirossian, F., \& Mezzetti, M.\ 1992, \apj, 396, 35 

\bibitem[Blanton et al. 2005]{blanton05} Blanton, M.~R., Lupton, 
R.~H., Schlegel, D.~J., Strauss, M.~A., Brinkmann, J., Fukugita, M., \& 
Loveday, J.\ 2005, \apj, 631, 208 


\bibitem[Blanton et al.  2003]{blanton03} Blanton, M.~R., et al.\ 
2003, \aj, 125, 2348

\bibitem[B{\"o}hringer et al.(2000)]{boringer00} B{\"o}hringer,
 H., et al.\ 2000, \apjs, 129, 435
 

\bibitem[Borgani et al. 1999]{borgani99} Borgani, S., Girardi, 
M., Carlberg, R.~G., Yee, H.~K.~C., \& Ellingson, E.\ 1999, \apj, 527, 561 

\bibitem[Butcher \& Oemler  1984]{butcher84} Butcher, H., \& 
Oemler, A.\ 1984, \apj, 285, 426 

\bibitem[Carlberg et al.  1997]{carlberg97} Carlberg, R.~G., Yee, 
H.~K.~C., \& Ellingson, E.\ 1997, \apj, 478, 462

\bibitem[Castro-Rodr{\'{\i}}guez et al.  2003]{castrorodriguez03} 
Castro-Rodr{\'{\i}}guez, N., Aguerri, J.~A.~L., Arnaboldi, M., Gerhard, O., 
Freeman, K.~C., Napolitano, N.~R., \& Capaccioli, M.\ 2003, \aap, 405, 803

\bibitem[Cole \& Lacey  1996]{cole96} Cole, S., \& Lacey, C.\ 
1996, \mnras, 281, 716 

\bibitem[Colless et al. 2001]{colless01} Colless, M., et al.\ 
2001, \mnras, 328, 1039

\bibitem[Curtis 1918]{curtis18} Curtis, H.~D.\ 1918, 
Publications of Lick Observatory, 13, 55

\bibitem[Danese et al.  1980]{danese80} Danese, L., de Zotti, 
G., \& di Tullio, G.\ 1980, \aap, 82, 322

\bibitem[den Hartog \& Katgert  1996]{denhartog96} den Hartog, R., 
\& Katgert, P.\ 1996, \mnras, 279, 349

\bibitem[De Propris et al.  2004]{depropris04} De Propris, R., et 
al.\ 2004, \mnras, 351, 125

\bibitem[De Propris et al. 2003]{depropris03} De Propris, R., et 
al.\ 2003, \mnras, 342, 725 

\bibitem[Dressler et al.  1997]{dressler97} Dressler, A., et al.\ 
1997, \apj, 490, 577

\bibitem[Dressler  1980]{dressler80} Dressler, A.\ 1980, \apj, 
236, 351 

\bibitem[Ebeling et al. 2000]{ebeling00} Ebeling, H., Edge,
 A.~C., Allen, S.~W., Crawford, C.~S., Fabian, A.~C., \& Huchra, J.~P.\
 2000, \mnras, 318, 333

\bibitem[Ebeling et al. 1998]{ebeling98} Ebeling, H., Edge,
 A.~C., Bohringer, H., Allen, S.~W., Crawford, C.~S., Fabian, A.~C., Voges,
 W., \& Huchra, J.~P.\ 1998, \mnras, 301, 881

\bibitem[Edge \& Stewart  1991]{edge91} Edge, A.~C., \& 
Stewart, G.~C.\ 1991, \mnras, 252, 414

\bibitem[Ellingson et al.  2001]{ellingson01} Ellingson, E., Lin, 
H., Yee, H.~K.~C., \& Carlberg, R.~G.\ 2001, \apj, 547, 609

\bibitem[Fadda et al.  1996]{fadda96} Fadda, D., Girardi, M., 
Giuricin, G., Mardirossian, F., \& Mezzetti, M.\ 1996, \apj, 473, 670

\bibitem[Fairley et al.  2002]{fairley02} Fairley, B.~W., Jones, 
L.~R., Wake, D.~A., Collins, C.~A., Burke, D.~J., Nichol, R.~C., \& Romer, 
A.~K.\ 2002, \mnras, 330, 755 

\bibitem[Fasano et al.  2000]{fasano00} Fasano, G., Poggianti, 
B.~M., Couch, W.~J., Bettoni, D., Kj{\ae}rgaard, P., \& Moles, M.\ 2000, 
\apj, 542, 673 



\bibitem[Fujita  1998]{fujita98} Fujita, Y.\ 1998, \apj, 509, 
587


\bibitem[Gallazzi et al. 2006]{gallazzi06} Gallazzi, A., Charlot, 
S., Brinchmann, J., \& White, S.~D.~M.\ 2006, \mnras, 370, 1106

\bibitem[Girardi \& Mezzetti  2001]{girardi01} Girardi, M., \& 
Mezzetti, M.\ 2001, \apj, 548, 79

\bibitem[Girardi et al.  1996]{girardi96} Girardi, M., Fadda, D., 
Giuricin, G., Mardirossian, F., Mezzetti, M., \& Biviano, A.\ 1996, \apj, 
457, 61 

\bibitem[G{\'o}mez et al.  2003]{gomez03} G{\'o}mez, P.~L., et 
al.\ 2003, \apj, 584, 210 

\bibitem[Goto  2005]{goto05} Goto, T.\ 2005, \mnras, 359, 1415

\bibitem[Goto et al.  2003]{goto03} Goto, T., et al.\ 2003, 
\pasj, 55, 739

\bibitem[Gott 1972]{gott72} Gott, J.~R.~I.\ 1972, \apj, 173, 
227 

\bibitem[Gunn \& Gott 1972 ]{gunn72} Gunn, J.~E., \& Gott, 
J.~R.~I.\ 1972, \apj, 176, 1 

\bibitem[Guti{\'e}rrez et al.  2004]{gutierrez04} Guti{\'e}rrez, 
C.~M., Trujillo, I., Aguerri, J.~A.~L., Graham, A.~W., \& Caon, N.\ 2004, 
\apj, 602, 664

\bibitem[Haines et al. 2006]{haines06} Haines, C.~P., La 
Barbera, F., Mercurio, A., Merluzzi, P., \& Busarello, G.\ 2006, \apjl, 
647, L21 


\bibitem[Hilton et al.  2005]{hilton05} Hilton, M., et al.\ 
2005, \mnras, 363, 661 

\bibitem[Hogg et al. 2004]{hogg04} Hogg, D.~W., et al.\ 2004, 
\apjl, 601, L29

   \bibitem[Hogg et al. 2001]{hogg01} Hogg, D.~W., Finkbeiner, 
D.~P., Schlegel, D.~J., \& Gunn, J.~E.\ 2001, \aj, 122, 2129

\bibitem[Hubble \& Humason(1931)]{hubble31} Hubble, E., \& 
Humason, M.~L.\ 1931, \apj, 74, 43


\bibitem[Katgert et al.  2004]{katgert04} Katgert, P., Biviano, 
A., \& Mazure, A.\ 2004, \apj, 600, 657

\bibitem[Katgert et al.  1998]{katgert98} Katgert, P., Mazure, 
A., den Hartog, R., Adami, C., Biviano, A., \& Perea, J.\ 1998, \aaps, 129, 
399  

\bibitem[Kenney et al. 2004]{kenney04} Kenney, J.~D.~P., van 
Gorkom, J.~H., \& Vollmer, B.\ 2004, \aj, 127, 3361 

\bibitem[Ledlow et al. 2003]{ledlow03} Ledlow, M.~J., Voges,
 W., Owen, F.~N., \& Burns, J.~O.\ 2003, \aj, 126, 2740


\bibitem[Lewis et al.  2002]{lewis02} Lewis, I., et al.\ 2002, 
\mnras, 334, 673

   \bibitem[Lupton et al. 2002]{lupton02} Lupton, R.~H., Ivezic, 
Z., Gunn, J.~E., Knapp, G., Strauss, M.~A., \& Yasuda, N.\ 2002, \procspie, 
4836, 350

\bibitem[Mahdavi \& Geller  2001]{mahdavi01} Mahdavi, A., \& 
Geller, M.~J.\ 2001, \apjl, 554, L129 

\bibitem[Mamon et al.  2004]{mamon04} Mamon, G.~A., Sanchis, 
T., Salvador-Sol{\'e}, E., \& Solanes, J.~M.\ 2004, \aap, 414, 445

\bibitem[Mamon  1992]{mamon92} Mamon, G.~A.\ 1992, \apjl, 401, 
L3 

\bibitem[Mastropietro et al. 2005]{mastroprieto05} Mastropietro, C., 
Moore, B., Mayer, L., Debattista, V.~P., Piffaretti, R., \& Stadel, J.\ 
2005, \mnras, 364, 607 

\bibitem[Margoniner et al.  2001]{margoniner01} Margoniner, V.~E., 
de Carvalho, R.~R., Gal, R.~R., \& Djorgovski, S.~G.\ 2001, \apjl, 548, 
L143 

\bibitem[Margoniner \& de Carvalho  2000]{margoniner00} Margoniner, 
V.~E., \& de Carvalho, R.~R.\ 2000, \aj, 119, 1562 

\bibitem[Melnick \& Sargent  1977]{melnick77} Melnick, J., \& 
Sargent, W.~L.~W.\ 1977, \apj, 215, 401

\bibitem[Metevier et al.  2000]{metevier00} Metevier, A.~J., 
Romer, A.~K., \& Ulmer, M.~P.\ 2000, \aj, 119, 1090

\bibitem[Miller et al. 2005]{miller05} Miller, C.~J., et al.\ 
2005, \aj, 130, 968 

\bibitem[Moore et al. 1996]{moore96} Moore, B., Katz, N., 
Lake, G., Dressler, A., \& Oemler, A.\ 1996, \nat, 379, 613 

\bibitem[Moss \& Dickens  1977]{moss77} Moss, C., \& Dickens, 
R.~J.\ 1977, \mnras, 178, 701

\bibitem[Mulchaey \& Zabludoff  1998]{mulchaey98} Mulchaey, J.~S., 
\& Zabludoff, A.~I.\ 1998, \apj, 496, 73

\bibitem[Muriel et al.  2002]{muriel02} Muriel, H., Quintana, 
H., Infante, L., Lambas, D.~G., \& Way, M.~J.\ 2002, \aj, 124, 1934 

\bibitem[Oemler  1974]{oemler74} Oemler, A.~J.\ 1974, \apj, 194, 1

\bibitem[Ortiz-Gil et al.  2004]{ortizgil04} Ortiz-Gil, A., Guzzo, 
L., Schuecker, P., B{\"o}hringer, H., \& Collins, C.~A.\ 2004, \mnras, 348, 
325 
 
   \bibitem[Pier et al. 2003]{pier03} Pier, J.~R., Munn, J.~A., 
Hindsley, R.~B., Hennessy, G.~S., Kent, S.~M., Lupton, R.~H., \& 
Ivezi{\'c}, {\v Z}.\ 2003, \aj, 125, 1559

\bibitem[Pisani 1996]{pisani96} Pisani, A.\ 1993, \mnras, 278, 697 

\bibitem[Poggianti et al. 2006]{poggianti06} Poggianti, B.~M., et 
al.\ 2006, \apj, 642, 188

\bibitem[Popesso et al. 2006]{popeso06} Popesso, P., Biviano, A., Bohringer, H. \& Romaniello, M., 2006, astro-ph/0606191

\bibitem[Quilis et al.  2000]{quilis00} Quilis, V., Moore, B., 
\& Bower, R.\ 2000, Science, 288, 1617 

\bibitem[Quintana \& Melnick  1982]{quintana82} Quintana, H., \& 
Melnick, J.\ 1982, \aj, 87, 972 

\bibitem[Rakos et al. 1997]{rakos97} Rakos, K.~D., Odell, 
A.~P., \& Schombert, J.~M.\ 1997, \apj, 490, 194

\bibitem[Rines \& Diaferio 2006]{rines06} Rines, K., \& 
Diaferio, A.\ 2006, \aj, 132, 1275 

\bibitem[Rines et al. 2003]{rines03} Rines, K., Geller, M.~J., 
Kurtz, M.~J., \& Diaferio, A.\ 2003, \aj, 126, 2152 


\bibitem[Schlegel et al.  1998]{schlegel98} Schlegel, D.~J., 
Finkbeiner, D.~P., \& Davis, M.\ 1998, \apj, 500, 525

\bibitem[Shimasaku et al.  2001]{shimasaku01} Shimasaku, K., et 
al.\ 2001, \aj, 122, 1238

\bibitem[Smail et al.  1998]{smail98} Smail, I., Edge, A.~C., 
Ellis, R.~S., \& Blandford, R.~D.\ 1998, \mnras, 293, 124 

\bibitem[Sodre et al.  1989]{sodre89} Sodre, L.~J., Capelato, 
H.~V., Steiner, J.~E., \& Mazure, A.\ 1989, \aj, 97, 1279

\bibitem[Solanes et al.  2001]{solanes01} Solanes, J.~M., 
Manrique, A., Garc{\'{\i}}a-G{\'o}mez, C., Gonz{\'a}lez-Casado, G., 
Giovanelli, R., \& Haynes, M.~P.\ 2001, \apj, 548, 97  

\bibitem[Stein  1997]{stein97} Stein, P.\ 1997, \aap, 317, 670 

\bibitem[Strateva et al.  2001]{strateva01} Strateva, I., et al.\ 
2001, \aj, 122, 1861

    \bibitem[Strauss et al. 2002]{strauss02} Strauss, M.~A., et al.\ 
2002, \aj, 124, 1810 

\bibitem[Tammann  1972]{tamman72} Tammann, G.~A.\ 1972, \aap, 
21, 355

\bibitem[Xue \& Wu  2000]{xue00} Xue, Y.-J., \& Wu, X.-P.\ 
2000, \apj, 538, 65 

\bibitem[Voges et al.(1999)]{voges99} Voges, W., et al.\ 1999,
 \aap, 349, 389
 


\bibitem[Yahil \& Vidal  1977]{yahil77} Yahil, A., \& Vidal, 
N.~V.\ 1977, \apj, 214, 347

    \bibitem[York et al. 2000]{york00} York, D.~G., et al.\ 2000, \aj, 120, 1579

\bibitem[Zabludoff et al. 1990]{zabludoff90} Zabludoff, A.~I., 
Huchra, J.~P., \& Geller, M.~J.\ 1990, \apjs, 74, 1

\bibitem[Zwicky et al.(1961)]{zwicky61} Zwicky, F., Herzog, E.,
 \& Wild, P.\ 1961, Pasadena: California Institute of Technology (CIT),
 |c1961,
\end{thebibliography}
\end{document}